\documentclass[11pt,fleqn]{article}
\usepackage{amssymb,latexsym,amsmath,amsfonts,graphicx, ebezier}

    \advance\textheight by \topskip

    \numberwithin{equation}{section}

\newcommand{\be}{\begin{equation}}
\newcommand{\ee}{\end{equation}}

\newcommand{\wt}{\widetilde}
\newcommand{\de}{\delta}

\newcommand{\al}{\alpha}

\newcommand{\si}{\sigma}
\newcommand{\Si}{\Sigma}
\newcommand{\om}{\omega}
\newcommand{\Om}{\Omega}
\newcommand{\lb}{\lambda}
\newcommand{\ze}{\zeta}

\renewcommand{\th}{\theta}

\newcommand{\ep}{\varepsilon }

    \def\tr{{\rm tr \,}}
    \def\Re{{\rm Re \,}}
    \def\Im{{\rm Im \,}}
    \def\Ai{{\rm Ai \,}}

    \def\bigO{{\cal O}}
    
    \def\Res{{\rm Res}}
    
    \def\P2n{{\rm P}_{{\rm II}}^{(n)}}

    \newtheorem{theorem}{Theorem}[section]
    \newtheorem{lemma}[theorem]{Lemma}

    \newtheorem{Definition}[theorem]{Definition}
    
    \newtheorem{Remark}[theorem]{Remark}
    \newenvironment{remark}{\begin{Remark}\rm}{\end{Remark}}
    \newtheorem{Example}[theorem]{Example}
    \newenvironment{example}{\begin{Example}\rm}{\end{Example}}
    \newtheorem{Assumptions}[theorem]{Assumptions}

    {\rm \trivlist \item[\hskip \labelsep{\bf Proof. }]}%
    {\hspace*{\fill}$\Box$\endtrivlist}
    {\rm \trivlist \item[\hskip \labelsep{\bf Proof}]}%
    {\hspace*{\fill}$\Box$\endtrivlist}

    \newcommand{\supp}{{\operatorname{supp}}}

    \DeclareMathOperator*{\Tr}{Tr}

\begin{document}
\title{Higher order analogues of the Tracy-Widom distribution and the Painlev\'e II hierarchy}
\author{T. Claeys, A. Its, and I. Krasovsky}

\maketitle

\begin{abstract}
We study Fredholm determinants related to a family of kernels which describe the edge eigenvalue behavior in unitary random matrix models with critical edge points. The kernels are natural higher order analogues of the Airy kernel and are built out of functions associated with
 the Painlev\'e I hierarchy. The Fredholm determinants related to those kernels are higher order generalizations of the Tracy-Widom distribution. We give an explicit expression for the determinants in terms of a distinguished smooth solution to the Painlev\'e II hierarchy. In addition we compute large gap asymptotics for the Fredholm determinants.

\end{abstract}

\section{Introduction}

In unitary random matrix ensembles with a probability measure of the form
\begin{equation}\label{random matrix model}
\frac{1}{Z_{n}}e^{-n\,\tr V(M)}dM,
\end{equation}
on the Hermitian $n\times n$ matrices, where $V$ is real analytic on $\mathbb R$ with sufficient growth at infinity, various correlation functions of eigenvalues
can be expressed in terms of the kernel
\begin{equation} \label{kernel}
    K_n(x,y)
        =\frac{e^{-\frac{n}{2}V(x)}
e^{-\frac{n}{2}V(y)}}{x-y}\frac{\kappa_{n-1}}{\kappa_{n}}
(p_n(x)p_{n-1}(y)-p_n(y)p_{n-1}(x)),
\end{equation} constructed from the polynomials
\[
    p_k(x)=\kappa_k x^k + \cdots,
    \qquad\qquad \mbox{$\kappa_k>0$,}
\]
orthonormal
with respect to the weight $e^{-nV}$ on $\mathbb R$.

\medskip

The limiting mean eigenvalue density is known, see e.g.\ \cite{Deift}, to be given as the density of an equilibrium measure $\mu_V$ minimizing the logarithmic energy
\begin{equation} \label{defIV}
    I_V(\mu)=\iint\log\frac{1}{|x-y|}d\mu(x)d\mu(y) +\int V(x)d\mu(x),
\end{equation}
among all probability measures $\mu$ on $\mathbb R$.
The equilibrium density depends on the potential $V$ but can in general be written in the form \cite{DKM}
\[\psi_V(x)=\frac{d\mu_V(x)}{dx}=\frac{1}{\pi}\sqrt{Q_V^-(x)},\qquad Q_V^-(x)=
\begin{cases}
-Q_V(x),& \mbox{if } Q_V(x)<0,\cr
0,& \mbox{otherwise},
\end{cases}\] where $Q_V$ is a real analytic function determined by $V$.
Generically $Q_V$ has simple zeros at the endpoints of $\supp\,\psi_V$, so that $\psi_V$ vanishes as a square root \cite{KM}. For special (non-regular or critical) $V$'s however,
the limiting mean eigenvalue density vanishes faster. In general,
$Q_V$ has a zero of order $4k+1$, $k=0,1,\dots$ at an endpoint of the support.

\begin{remark}\label{remark: endpoints}
$Q_V$ cannot have zeros of order $4k+3$. These endpoint behaviors would contradict variational conditions that follow from the minimization property of the equilibrium measure \cite{DKMVZ2}.
\end{remark}

\begin{example}
The simplest non-regular case $k=1$ is realized, e.g., for a critical quartic potential
    \begin{equation}\label{V}
        V(x)=\frac{1}{20}x^4-\frac{4}{15}x^3+\frac{1}{5}x^2+\frac{8}{5}x.
    \end{equation}
    Here the limiting mean eigenvalue density is supported on $[-2,2]$ and given by
    \begin{equation}
        \psi_V(x) =
        \frac{1}{10\pi}(x+2)^{1/2}(2-x)^{5/2}\chi_{[-2,2]}(x).
    \end{equation}
It is easy to verify (\ref{V}) by substituting this $\psi_V(x)$ into the
variational conditions. In fact, it is much easier to recover $V$ starting from a given limiting mean eigenvalue density than to find the limiting mean eigenvalue density corresponding to a given potential $V$ directly.

In order to construct polynomial potentials $V$ for which $k\ge 1$, it is necessary that the degree of $V$ is at least $2k+2$.
\end{example}

In the generic case where $k=0$ for the rightmost endpoint $b=b_V$
of $\supp\,\psi_V$,
it follows from the results of
Deift {\it et al} \cite{DKMVZ2,DG} that, for any fixed $u$ and $v$,
\begin{equation}\label{Airykernel}
\lim_{n\to\infty}\frac{1}{cn^{2/3}}K_n(b+\frac{u}{cn^{2/3}},
b+\frac{v}{cn^{2/3}})=K^{(0)}(u,v),
\end{equation}
where
\begin{equation}\label{kernel1}
K^{(0)}(u,v)=\frac{\Ai(u)\Ai'(v)-\Ai(v)\Ai'(u)}{u-v}
\end{equation}
is the Airy kernel, and $c=c_V$ is a constant depending on $V$.
Let $\lambda_n$ be the largest eigenvalue of a random matrix $M$.
It was proved in many cases \cite{TW,DG} and is believed to hold for $k=0$
in general that the limiting distribution of $\lambda_n$ is given by
\begin{equation}\label{TW0}
\lim_{n\to\infty}{\rm Prob}\left(cn^{2/3}(\lambda_n-b)<s\right)=\det(I-K_s^{(0)}),
\end{equation}
where
$K_s^{(0)}$ is the Airy-kernel
trace-class operator acting on $L^2(s,\infty)$.

The function at the r.h.s. of (\ref{TW0}) is known as the Tracy-Widom distribution.
Tracy and Widom \cite{TW} discovered a representation
\begin{equation}\label{TW}
\det(I-K_s^{(0)})=\exp\left(-\int_s^{+\infty}(y-s)q_0^2(y)dy\right)
\end{equation}
in terms of the Hastings-McLeod solution $q_0$ of the Painlev\'e II equation
\begin{equation}\label{PII0}
q_{xx}=xq+2q^3
\end{equation}
characterized by the asymptotic behavior
\begin{align}&\label{HastingsMcLeod}
q_0(x)\sim \Ai(x), &\mbox{ as $x\to +\infty$,}\\
&\label{HastingsMcLeod2}q_0(x)= \sqrt\frac{-x}{2}\left(1 + \frac{1}{8x^{3}} +
O\left(x^{-6}\right)\right), &\mbox{ as $x\to -\infty$.}
\end{align}
The Tracy-Widom function $\det(I-K_s^{(0)})$ does not only describe the largest eigenvalue distribution in random matrix ensembles, but appears also in several combinatorial models, for example, it provides the distribution of the longest increasing subsequence of random permutations \cite{BDJ}.

\medskip

In the general case of arbitrary $k$, a family of limiting kernels $K^{(k)}(u,v)$
appears in place of the Airy kernel.
Consider a potential $V(x)$ such that $Q_V(x)$ has a
zero of order $4k+1$, $k=0,1,\dots$ at the rightmost endpoint of $\supp\,\psi_V$.
Furthermore, to obtain more general results,
consider for suitable $V_j(x)$ the deformation
\[
\wt V(x)=V(x)+\sum_{j=0}^{2k-1}T_j V_j(x),
\]
where $T_j$ are constants. The natural analogues of (\ref{Airykernel}) are
double-scaling limits where $n\to\infty$ and simultaneously $T_j\to 0$
at an appropriate rate in $n$ characterized by parameters $t_j$ (in the simplest
case we can take all $T_j=t_j=0$). We refer to \cite{CV2} for more details about those double-scaling limits in the case $k=1$.
In these limits we expect the following expressions for the kernel
(corresponding to $\wt V(x)$) and
the distribution of the largest eigenvalue, according to conjectures
in the physics
literature \cite{BB,BMP}:
\be\label{kernel20}
\lim_{n\to\infty}\frac{1}{cn^{2/(4k+3)}}K_n(b+\frac{u}{cn^{2/(4k+3)}},
b+\frac{v}{cn^{2/(4k+3)}})=K^{(k)}(u,v;t_0,\dots,t_{2k-1}),
\ee
for certain constants $b$ and $c$ and any fixed $u$, $v$. Here
\begin{equation}\label{kernel2}
K^{(k)}(u,v;t_0, \ldots , t_{2k-1})=\frac{\Phi_1^{(2k)}(u)\Phi_2^{(2k)}(v)-\Phi_1^{(2k)}(v)\Phi_2^{(2k)}(u)}{-2\pi i(u-v)},
\end{equation}
where the functions
$\Phi_j^{(2k)}(w)=\Phi_j^{(2k)}(w;t_0,\ldots,t_{2k-1})$ are described below.
Moreover,
\begin{multline}\label{distr}
\lim_{n\to\infty}{\rm Prob}\left(cn^{2/(4k+3)}
(\lambda_n-b)<s\right)\\=
\lim_{n\to\infty}\det(I-K_n\chi_{(b+s/[cn^{2/(4k+3)}],+\infty)})=
\det(I-K_s^{(k)}),
\end{multline}
where $K_n\chi_{(a,b)}$ is the operator with kernel $K_n$ acting on $L^2(a,b)$, and
$K_s^{(k)}$ is the trace-class operator with kernel (\ref{kernel2})
acting on $L^2(s,\infty)$.

Note that (\ref{kernel20}) for the case $k=1$ was proved in \cite{CV2}.
Our goal in this paper is not to prove (\ref{kernel20}) and (\ref{distr}).
Rather, we study the properties of $\det(I-K_s^{(k)})$.

\medskip

The functions
$\Phi_1^{(2k)}=\Phi_1^{(2k)}(\zeta;t_0, \ldots , t_{2k-1})$ and $\Phi_2^{(2k)}=\Phi_2^{(2k)}(\zeta;t_0, \ldots , t_{2k-1})$ appear as solutions of the Lax
pair associated with a distinguished solution to the $2k$-th member of the Painlev\'e I hierarchy \cite{BB, BMP, CV2}. They are most easily characterized in terms of the following Riemann-Hilbert (RH) problem.

\subsubsection*{RH problem for $\Phi$}

\begin{figure}[t]
    \begin{center}
    \setlength{\unitlength}{1mm}
    \begin{picture}(95,47)(0,2)
        \put(30,38){\small $\Gamma_2$}
        \put(13,27){\small $\Gamma_3$}
        \put(30,11){\small $\Gamma_4$}
        \put(75,27){\small $\Gamma_1$}

        \put(50,25){\thicklines\circle*{.9}}
        \put(51,21){\small 0}

        \put(50,25){\line(-2,1){35}} \put(36,32){\thicklines\vector(2,-1){.0001}}
        \put(50,25){\line(-2,-1){35}} \put(36,18){\thicklines\vector(2,1){.0001}}
        \put(50,25){\line(-1,0){40}} \put(30,25){\thicklines\vector(1,0){.0001}}
        \put(50,25){\line(1,0){35}} \put(70,25){\thicklines\vector(1,0){.0001}}
    \end{picture}
    \caption{Contour for the $\Phi$-RH problem. }
        \label{figure: gamma}
    \end{center}
\end{figure}
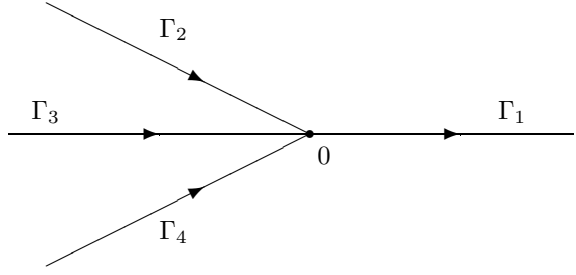

\begin{itemize}
    \item[(a)] $\Phi=\Phi^{(2k)}:\mathbb C\setminus \Gamma\to \mathbb C^{2\times 2}$ is analytic, with \[\Gamma=\cup_{j=1}^4\Gamma_j\cup\{0\}, \qquad \Gamma_1=\mathbb R^+,\quad\Gamma_3=\mathbb R^-, \quad \Gamma_2=e^{\frac{-i\pi}{4k+3}}\mathbb R^-, \quad \Gamma_4=e^{\frac{i\pi}{4k+3}}\mathbb R^-,\] oriented as in Figure \ref{figure: gamma}.
    \item[(b)] $\Phi$ has $L^2$ boundary values $\Phi_+$ as $\ze$ approaches
$\Gamma$ from the left (w.r.t. the direction shown by an arrow), and $\Phi_-$,
from the right. They are related by the jump conditions
    \begin{align}
        \label{RHP Psik: b1}
        &\Phi_+(\zeta)=\Phi_-(\zeta)
        \begin{pmatrix}
            0 & 1 \\
            -1 & 0
        \end{pmatrix},& \mbox{for $\zeta\in\Gamma_3$,} \\[1ex]
        \label{RHP Psik: b2}
        &\Phi_+(\zeta)=\Phi_-(\zeta)
        \begin{pmatrix}
            1 & 1 \\
            0 & 1
        \end{pmatrix},& \mbox{for $\zeta\in\Gamma_1$,} \\[1ex]
        \label{RHP Psik: b3}
        &\Phi_+(\zeta)=\Phi_-(\zeta)
        \begin{pmatrix}
            1 & 0 \\
            1 & 1
        \end{pmatrix},& \mbox{for $\zeta\in\Gamma_2\cup \Gamma_4$.}
    \end{align}
    \item[(c)] $\Phi$ has the following behavior as $\zeta\to\infty$:
    \begin{equation}\label{RHP Psik: c}
        \Phi(\zeta)=\zeta^{-\frac{1}{4}\sigma_3}N\left(I+\Phi_\infty\zeta^{-1/2}+
    \bigO(\zeta^{-1})\right)
        e^{-\theta(\zeta)\sigma_3},
    \end{equation}
    where $\Phi_\infty$ is independent of $\zeta$ (its explicit expression will not be
    important below), $\sigma_3$ is the Pauli matrix $\begin{pmatrix}1&0\\0&-1\end{pmatrix}$, $N$ is given by
    \begin{equation}\label{def: N0}
        N=\frac{1}{\sqrt 2}
        \begin{pmatrix}
             1 & 1 \\
             -1 & 1
        \end{pmatrix}e^{-\frac{1}{4}\pi i\sigma_3},
    \end{equation}
    and
    \begin{equation}\label{def theta0}
        \theta(\zeta;t_0,\ldots,t_{2k-1})=\frac{2}{4k+3}\zeta^{\frac{4k+3}{2}}-2\sum_{j=0}^{2k-1}\frac{
        (-1)^{j}t_{j}}{2j+1}\zeta^{\frac{2j+1}{2}},
    \end{equation}
    where the fractional powers denote (as usual throughout this paper) the principal branches analytic for $\zeta\in \mathbb C\setminus (-\infty,0]$ and positive for $\zeta>0$.
    \item[(d)] $\Phi$ is bounded near $0$.
\end{itemize}
The functions $\Phi_1=\Phi_1^{(2k)}$ and $\Phi_2=\Phi_2^{(2k)}$ appearing in (\ref{kernel2}) are the analytic extensions of the functions $\Phi_{11}$ and $\Phi_{21}$
from the sector in between $\Gamma_1$ and $\Gamma_2$ to the entire complex plane.
(One verifies the analyticity by multiplying
the jump matrices: see Remark \ref{remark: first column}.)
\medskip

A solution $\Phi$ to this RH problem satisfies a linear system of the form
\be\label{system}
\Phi_\zeta'=A\Phi,\qquad \Phi_{t_j}'=B_j\Phi,
\ee
where $A$ is a polynomial in $\zeta$ of degree $2k+1$, and $B_j$ is a polynomial in $\zeta$ of degree $j+1$.
The compatibility conditions for the system (\ref{system}) yield
the ${\rm P}_{\rm I}^{(2k)}$ equation, the Painlev\'e I equation of order $4k$. The most general form of the RH problem for ${\rm P}_{\rm I}^{(2k)}$ involves a contour consisting of $4k+4$ rays instead of $4$. Our RH problem is the one corresponding to one particular solution of ${\rm P}_{\rm I}^{(2k)}$, which was introduced by Br\'ezin, Marinari, and Parisi in \cite{BMP}.
The precise structure of the Painlev\'e I hierarchy and the relation of the functions $\Phi_1$ and $\Phi_2$ to the ${\rm P}_{\rm I}^{(2k)}$ equation is not important for us, but we refer the interested reader to \cite{KS, Kapaev, Kapaev2}. The Fredholm determinant $\det(I-K_s^{(k)})$ is the object we want to study, and we only need to know the RH characterization of $\Phi_1$ and $\Phi_2$ for that purpose.

\begin{remark}\label{remark poles}
In the case $k=1$, the existence of a RH solution $\Phi$ was proved in \cite[Lemma 2.3]{CV1} for all real values
 of $t_0, t_1$. The proof of this fact is however valid for arbitrary $k$ and arbitrary real values of the parameters $t_j$. It is important to stress that the proof does not hold for odd members of the Painlev\'e I hierarchy, in other words for the function $\Phi^{(2k+1)}$. Moreover, the RH problem for $\Phi^{(1)}$ is not solvable for every real $t_0$, because this would imply \cite{DK, CV1} the existence of a real-valued Painlev\'e I solution without poles on the real line, which contradicts \cite{JK}. However, as a consequence of Remark \ref{remark: endpoints}, only the kernels generated by the even members of the Painlev\'e I hierarchy appear in unitary random matrix ensembles of the form (\ref{random matrix model}).
\end{remark}

\begin{remark}\label{remark uniqueness}
It should be noted that the RH problem for $\Phi^{(2k)}$ is not uniquely
solvable: there is a family of solutions of the form
$\begin{pmatrix}1&0\\ \om&1\end{pmatrix}\Phi^{(2k)}(\zeta)$, where $\om$ can
depend on $t_0, \ldots , t_{2k-1}$ but not on $\zeta$. Indeed it is
obvious that the left multiplication by a constant matrix (independent
of $\zeta$) does not modify the jump conditions. If the matrix is
lower-triangular with $1$ on both diagonal entries, neither the
asymptotic condition (\ref{RHP Psik: c}) is violated. One
verifies as in \cite{IKO} that, up to the left multiplication by such a matrix,
the RH problem is uniquely
solvable. However, the kernel (\ref{kernel2}) is independent of the
choice of a solution and thus is well-defined.
\end{remark}

It follows from
(\ref{kernel2}) and (\ref{RHP Psik: c}) that
\begin{equation}\label{kernel 1+}K^{(k)}(u,v;t_0, \ldots , t_{2k-1})=
\bigO(e^{-cu^{\frac{4k+3}{2}}-cv^{\frac{4k+3}{2}}}),\qquad\mbox{ as $u,v\to +\infty$},\end{equation} for some constant $c>0$. Consequently,
the asymptotics of $\ln\det(I-K_s^{(k)})$ can be obtained by a standard
series expansion. In particular,
\begin{equation}\label{kernel +infty}
\ln\det(I-K_s^{(k)})=\bigO(e^{-cs^{\frac{4k+3}{2}}}), \qquad
\mbox{ as $s\to +\infty$}.
\end{equation}
The question of the asymptotic behavior of $\det(I-K_s^{(k)})$ as $s\to -\infty$
is much more subtle and will be addressed below.

\subsection{Statement of results}
In the present paper,
we will answer the following questions.
What are the ``large gap'' asymptotics for the Fredholm determinant
$\det(I-K_s^{(k)})$ as $s\to -\infty$?
Can we find an identity for $\det(I-K_s^{(k)})$ which generalizes
the Tracy-Widom formula (\ref{TW}) to general $k$?
What is the analogue of the Hastings-McLeod solution of Painlev\'e II
for general $k$, and what are its properties?

\subsubsection{Large gap asymptotics}
The first goal of this paper is to find large gap asymptotics for the Fredholm determinant $\det(I-K_s^{(k)})$ as $s\to -\infty$. Those asymptotics are known for
the Airy kernel, i.e. in the case $k=0$, namely
\begin{equation}\label{exp Airy}
\ln\det(I-K_s^{(0)})=-\frac{|s|^3}{12}-\frac{1}{8}\ln |s| +\chi +\bigO(|s|^{-3/2}),
\end{equation}
where
\begin{equation}\label{chi}
\chi=\frac{1}{24}\ln 2 +\zeta'(-1),
\end{equation}
and $\zeta(s)$ is the Riemann zeta-function. The first two terms
at the r.h.s. of (\ref{exp Airy}) follow easily from (\ref{TW}) and
(\ref{HastingsMcLeod2}). The expansion of the derivative of (\ref{exp Airy})
was obtained by Tracy and Widom who also conjectured the value
(\ref{chi}) for the constant $\chi$ \cite{TW}.
A full proof of (\ref{exp Airy}), (\ref{chi}) was given recently in \cite{DIK} and
another proof followed shortly in \cite{BBD}.

We obtain

\begin{theorem}\label{theorem: large gap}
Let $K^{(k)}$ be the kernel defined in (\ref{kernel2}) for arbitrary $k=0,1,2,\dots$, depending on parameters $t_0, \ldots , t_{2k-1}$, and write $K_s^{(k)}$ for the trace-class operator with kernel $K^{(k)}$ acting on $L^2(s,+\infty)$.
The asymptotic expansion for the Fredholm determinant $\det(I-K_s^{(k)})$
as $s\to -\infty$ is given by the formula
\begin{equation}\label{large gap expansion}
\frac{d}{ds}\ln \det(I-K_s^{(k)})=\frac{1}{4}a_0^2(s)|s|^{4k+2}+\frac{3a_1(s)}{16a_0(s)|s|}+\bigO(|s|^{-\frac{4k+5}{2}}).
\end{equation}
Here $a_0(s,t_0, \ldots , t_{2k-1})$ and $a_1(s, t_0, \ldots, t_{2k-1})$ are defined as follows:
\begin{multline}
\label{def aj1-0}
a_j(s)={1\over\Gamma(j+\frac{3}{2})}
\left(
\frac{\Gamma(2k +\frac{3}{2})}{\Gamma(2k+2-j)}\right.\\ \left.
+\sum_{m=2}^{2k+1-j}t_{2k+1-m}\frac{\Gamma(2k+\frac{3}{2}-m)}{\Gamma(2k+2-j-m)}
|s|^{-m}\right),
\end{multline}
where the sum vanishes if $j=2k$, $j=2k+1$, and
$\Gamma(x)$ is Euler's $\Gamma$-function.
\end{theorem}

\begin{remark}\label{remcon}
Expanding the right-hand side of (\ref{large gap expansion}) in negative powers of $|s|$ and integrating, one easily finds asymptotics for $\det(I-K_s^{(k)})$ itself except for the constant of integration $\chi^{(k)}$.
In particular, to the leading order
\be
\ln \det(I-K_s^{(k)})=
-\left(\frac{(4k+1)!}{2^{4k+1}(2k)!(2k+1)!}\right)^2
\frac{|s|^{4k+3}}{4k+3}\left[1+o(1)\right],
\ee
which was conjectured in \cite{BH, CET}.
The constant $\chi^{(k)}$ has a representation, as in the Airy case $k=0$,
in terms of a solution of a Painlev\'e equation and also in terms of a limit
of multiple integrals (see Section \ref{const}).
For $k=0$, the relevant multiple integral can be reduced to
an explicitly computable Selberg integral, which allowed the authors in
\cite{DIK} to obtain the simple expression (\ref{chi}).
\end{remark}

\begin{example}
For $k=0$, we have $a_0(s)=1$ and $a_1(s)=\frac{2}{3}$, so that (\ref{large gap expansion}) takes the form
\[\frac{d}{ds}\ln \det(I-K_s^{(0)})=\frac{1}{4}|s|^{2}+\frac{1}{8|s|}+\bigO(|s|^{-\frac{5}{2}}).\] Integrating this expression gives (\ref{exp Airy})
without fixing the constant $\chi$.

\medskip

For $k=1$,
\[a_0(s;t_0,t_1)=\frac{5}{8}+t_1|s|^{-2}+2t_0|s|^{-3}, \qquad a_1(s;t_0,t_1)=\frac{5}{4}+\frac{2}{3}t_1|s|^{-2}.\]
Theorem \ref{theorem: large gap} then gives
\[\frac{d}{ds}\ln \det(I-K_s^{(1)})=\frac{5^2}{2^8}|s|^6+\frac{5}{16}t_1|s|^4+\frac{5}{8}t_0|s|^3+\frac{1}{4}t_1^2|s|^2+t_0t_1|s|
+t_0^2+\frac{3}{8|s|}+\bigO(|s|^{-3}),\]
which integrates to
\begin{multline}\label{k1}
\ln \det(I-K_s^{(1)})=-\frac{5^2}{2^8\cdot 7}|s|^7-\frac{1}{16}t_1|s|^5-\frac{5}{32}t_0|s|^4-\frac{1}{12}t_1^2|s|^3-\frac{1}{2}t_0t_1|s|^2
-t_0^2|s| \\ +\chi^{(1)}-\frac{3}{8}\ln|s|+\bigO(|s|^{-2}),\end{multline}
where $\chi^{(1)}$ is the constant of integration. In Section \ref{const},
we discuss 2 representations for this constant mentioned in Remark \ref{remcon}.
\end{example}

\subsubsection{The Painlev\'e II hierarchy}

On our way to the higher order analogues of the Tracy-Widom formula, let us first describe the Painlev\'e II hierarchy and in particular one distinguished solution to it, which is a natural higher order analogue of the Hastings-McLeod solution to the Painlev\'e II equation.

\medskip

The $n$-th member of the Painlev\'e II hierarchy, the ${\rm P}_{{\rm II}}^{(n)}$ equation, is the following differential equation of order $2n$ for $q=q(x;\tau_1, \ldots, \tau_{n-1})$ \cite{MazzoccoMo, Clarkson,JM, KS}:
\begin{equation}\label{PIIn}
\left(\frac{d}{dx}+2q\right)\mathcal L_n[q_x-q^2]+\sum_{\ell=1}^{n-1}\tau_{\ell} \left(\frac{d}{dx}+2q\right)\mathcal L_\ell[q_x-q^2]=xq-\alpha, \qquad n\geq 1,
\end{equation}
where the operator $\mathcal L_n$ is defined by the Lenard recursion
relation
\begin{equation}\label{def L}
\frac{d}{dx}\mathcal L_{j+1}f=\left(\frac{d^3}{dx^3}+4f\frac{d}{dx}+2f_x\right)\mathcal L_j f, \qquad \mathcal L_0 f=\frac{1}{2},\qquad \mathcal L_j 0=0,\quad j\ge 1.
\end{equation}
In particular, the first members of the hierarchy (we write $q_x^{(j)}$ for the $j$-th derivative of $q$ with respect to $x$) are given by the formulas:
\begin{align}
&{\rm P}_{{\rm II}}^{(1)}: && q_{xx}-2q^3=xq-\alpha,\\
&{\rm P}_{{\rm II}}^{(2)}: && (q_{x}^{(4)}-10 qq_x^2-10q^2q_{xx}+6q^5)+\tau_1(q_{xx}-2q^3)=xq-\alpha,\\
&{\rm P}_{{\rm II}}^{(3)}: && (q_x^{(6)}-14q^2q_x^{(4)}-56qq_xq_x^{(3)}-70(q_x)^2q_{xx}-42q(q_{xx})^2\nonumber \\
&\ &&\qquad +70q^4q_{xx}+140q^3q_x^2-20q^7) +\tau_2(q_{x}^{(4)}-10 qq_x^2-10q^2q_{xx}+6q^5)\nonumber \\
&\ &&\qquad\qquad\qquad\qquad\qquad\qquad\qquad\qquad+\tau_1(q_{xx}-2q^3)=xq-\alpha.
\end{align}
We are interested in the odd members of
the above hierarchy, with the value of the parameter
$\alpha=\frac{1}{2}$. One distinguished solution of the ${\rm P}_{{\rm II}}^{(n)}$ equation attracts our attention. It has the following properties.

\begin{theorem}\label{theorem: q}
Fix $\tau_1, \ldots, \tau_{n-1}\in\mathbb{R}$.
For $n$ odd and $\alpha=\frac{1}{2}$, there exists a real solution $q=q(x;\tau_1, \ldots, \tau_{n-1})$ of the ${\rm P}_{{\rm II}}^{(n)}$ equation (\ref{PIIn}) which has no poles for the real values of $x$, and the following asymptotic behavior as $x\to \pm\infty$:
\begin{align}\label{asymptotics q+}
&q(x)= \frac{1}{2x}+\bigO(x^{-\frac{4n+1}{2n}}), &\qquad \mbox{ as $x\to +\infty$},\\
&\label{asymptotics q-}q(x)=
\left(\frac{n!^2}{(2n)!}|x|\right)^{\frac{1}{2n}}+\bigO(|x|^{-1}),
&\qquad \mbox{ as $x\to -\infty$}.
\end{align}
\end{theorem}

\begin{remark}
For $n=1$, (\ref{asymptotics q+}) and (\ref{asymptotics q-}) together fix the solution to be the Hastings-McLeod solution of Painlev\'e II \cite{HastingsMcLeod}. Note that the function $q$ differs from the function $u$ given by (\ref{PII0})-(\ref{HastingsMcLeod2}), which is the Hastings-McLeod solution corresponding to the value $\alpha=0$, while $q$ corresponds to $\alpha=\frac{1}{2}$. We will explain below why the case $\alpha=\frac{1}{2}$, and not $\alpha=0$, is relevant for the Fredholm determinant if $k\geq 1$.
For general odd $n$, although the $\P2n$ equation is of order $2n$, it is our belief that (\ref{asymptotics q+}) and (\ref{asymptotics q-}) together still (as for $n=1$) determine the real solution $q$ uniquely. We do not have a proof of this fact, but we will provide some supporting heuristic arguments later on.
\end{remark}

\begin{remark}The asymptotics (\ref{asymptotics q+}) at $+\infty$ indicate that the terms at the right-hand side of (\ref{PIIn}) are dominant compared to those at the left-hand side, and that there is a balance between the terms $xq$ and $\alpha=\frac{1}{2}$. There is a whole family of solutions sharing the local asymptotic behavior (\ref{asymptotics q+}), see \cite{JM}.
The asymptotics (\ref{asymptotics q-}) at $-\infty$ indicate a balance of the $q^{2n+1}$-term at the left-hand side with the term $xq$ at the right-hand side. Again the existence of solutions with this local behavior was shown in \cite{JM}.
\end{remark}

\begin{remark}
It is a well-known fact that there is a one-to-one map between the solutions of a Painlev\'e equation and a so-called monodromy surface \cite{FIKN}.
The solution considered in Theorem \ref{theorem: q} is mapped to the point on the monodromy surface corresponding to the Stokes multipliers
\begin{align*}&s_2=s_3=\cdots =s_{2n}=0,\\
&s_1=s_{2n+1}=-i.
\end{align*} We will characterize this Painlev\'e transcendent
in Section \ref{section Painleve} in terms of a RH problem.
\end{remark}

\subsubsection{Painlev\'e expression for the Fredholm determinants}

Let $n=2k+1$. Define $b_j=b_j(s;t_0, \ldots , t_{2k-1})$ for $j=0, \ldots ,2k$ slightly modified compared to $a_j$
(\ref{def aj1-0}) by
\begin{multline}
\label{def bj1-0}
b_j(s)=\frac{\Gamma(2k +\frac{3}{2})}{\Gamma(j+\frac{3}{2})\Gamma(2k+2-j)}s^{2k+1-j}\\
-\sum_{\ell=j}^{2k-1}(-1)^\ell t_\ell \frac{\Gamma(\ell +\frac{1}{2})}{\Gamma(j+\frac{3}{2})\Gamma(\ell - j+1)}s^{\ell-j},
\end{multline}
and let $x=x(s;t_0, \ldots , t_{2k-1})$ and $\tau_j=\tau_j(s;t_0, \ldots , t_{2k-1})$ be polynomials in $s$ defined by
\begin{align}
&\label{x-0}x(s)=-2^{\frac{4k+1}{4k+3}}b_0(s;t_0, \ldots, t_{2k-1}),\\
&\label{ttau-0}\tau_{j}(s)=(2j+1)2^{\frac{4(k-j)+1}{4k+3}}
b_j(s;t_0, \ldots, t_{2k-1}), &j=1, \ldots , 2k.
\end{align}
For $k=0$ all $t_j$'s, $\tau_j$'s, and the sum in (\ref{def bj1-0}) vanish.

We prove

\begin{theorem}\label{theorem: TW}
Let $K^{(k)}$ be the kernel defined in (\ref{kernel2}) for arbitrary $k=0,1,\dots$, and write $K_s^{(k)}$ for the trace-class operator with kernel $K^{(k)}$ acting on $L^2(s,+\infty)$. Then the following identities hold:
\begin{equation}\label{Fredholmidentitya}
\frac{d}{ds}\ln\det(I-K_{s}^{(k)})=Q[x(s);\tau_1(s), \ldots , \tau_{2k}(s)],
\end{equation}
where $x(s), \tau_1(s), \ldots , \tau_{2k}(s)$ are given by (\ref{x-0})-(\ref{ttau-0}). Furthermore, for any fixed $\tau_1,\ldots ,\tau_{2k}$,
\begin{equation}\label{Qinfty}
Q(x;\tau_1, \ldots, \tau_{2k})=\int_{-\infty}^x
u(\xi;\tau_1, \ldots, \tau_{2k})^2d\xi
\end{equation}
with $u(x)=u(x;\tau_1, \ldots, \tau_{2k})$ given by
\begin{equation}\label{def u0}
u(x)=2^{-\frac{4k+1}{4k+3}}\exp\left\{-\int_2^{+\infty}\left(q(\xi)-\frac{1}{2\xi}\right)d\xi\right\}\cdot \exp\left\{\int_2^xq(\xi)d\xi\right\},
\end{equation}
where $q(x;\tau_1, \ldots , \tau_{2k})$ is the solution of the ${\rm P}_{{\rm II}}^{(2k+1)}$ equation characterized by the Stokes multipliers \[s_2=s_3=\cdots
=s_{4k+2}=0, \qquad s_1=s_{4k+3}=-i.\]
This solution satisfies the conditions of Theorem \ref{theorem: q}.
\end{theorem}

\begin{remark}
By (\ref{x-0}), the limit $x\to\pm\infty$ corresponds to $s\to\mp\infty$.
\end{remark}

\begin{remark}
The function $u(x)$ with fixed $\tau_1,\ldots ,\tau_{2k}$
can also be characterized as
the solution of the linear second order differential equation
\begin{equation}\label{DE2}
u''(x)=[q_x(x)+q(x)^2] u(x),
\end{equation}
with boundary conditions given by
\begin{align}
&\label{boundary1}u(x)=
2^{-\frac{4k+1}{4k+3}}\sqrt{\frac{x}{2}}(1+o(1)), &\mbox{ as $x\to +\infty$},\\
&\label{boundary2}u(x)=o(1), &\mbox{ as $x\to -\infty$.}
\end{align}
This allows us to check that our result reproduces
the Tracy-Widom formula (\ref{TW}) for $k=0$.
In this case $x=-2^{1/3}s$, and furthermore,
there exists a Backlund transformation relating $q$ (solving Painlev\'e II with $\alpha=\frac{1}{2}$) with the usual Hastings-McLeod solution $q_0$ (corresponding to Painlev\'e II (\ref{PII0}) with $\alpha=0$). Indeed we have
\cite{BBIK, Clarkson}
\be\label{Bt}
2^{-4/3}(x+2q^2(x)+2q_x(x))=q_0(-2^{-1/3}x)^2.
\ee
Equation (\ref{DE2}) thus becomes
\begin{equation}\label{DE2b}
u''(x)=[2^{1/3}q_0(-2^{-1/3}x)^2-\frac{x}{2}] u(x).
\end{equation}
The substitution $u(x)=2^{-1/6}q_0(-2^{-1/3}x)=2^{-1/6}q_0(s)$ reduces
this equation to the Painlev\'e II equation (\ref{PII0}) for $q_0(s)$ w.r.t.
the variable $s$.
Moreover, we conclude from the boundary conditions (\ref{boundary1}), (\ref{boundary2}) that  $q_0(s)$ is the Hastings-McLeod solution.
Now taking one more $s$-derivative of (\ref{Fredholmidentitya}) gives
\[\frac{d^2}{ds^2}\ln\det(I-K_{s}^{(k)})=-q_0(s)^2,\] which is equivalent to the Tracy-Widom formula (\ref{TW}).
\end{remark}

\begin{remark}
The function $q_x(x)+q(x)^2$ appearing in (\ref{DE2}) is a solution of
a differential equation
which is also known as the $(2k+1)$-th member of the Painlev\'e XXXIV hierarchy, and which is equivalent to the $(2k+1)$-th member of the Painlev\'e II hierarchy, see \cite{Clarkson} and also \cite{BBIK, IKO}. Since the Painlev\'e II hierarchy is more standard and better known, we prefer to state our results
in terms of it.
For $k=0$, the function $q_x+q^2$ can, as explained above, be expressed in terms of the Hastings-McLeod solution $q_0$ for $\alpha=0$, relying on the Backlund transformation (\ref{Bt}) which relates Painlev\'e II for $\alpha=1/2$ with $\alpha=0$ via the Painlev\'e XXXIV equation \cite{Clarkson}.
For $k>0$, a transformation for the Painlev\'e II hierarchy relating the values $\alpha=\frac{1}{2}$ and $\alpha=0$ is, to the best of our knowledge,
not present in the literature.
\end{remark}

\subsubsection*{Outline for the rest of the paper}
The proofs of Theorems \ref{theorem: large gap} and \ref{theorem: TW} are based on a RH representation for the logarithmic derivative of the Fredholm determinants $\det(I-K_{s}^{(k)})$. We present this identity in Section \ref{section fredholm}. In Section \ref{section large gap}, we obtain asymptotics for $\frac{d}{ds}\ln\det(I-K_{s}^{(k)})$ as $s\to -\infty$ by applying the steepest descent method of Deift and Zhou to the associated RH problem. Section \ref{section q} is devoted to the proof of Theorem \ref{theorem: q}, which relies on the study of a RH problem for the Painlev\'e II hierarchy. In Section \ref{section Painleve}, we identify the RH problem associated to the Fredholm determinant with the one related to the Painlev\'e II hierarchy. This enables us to give higher order analogues of the Tracy-Widom formula and thus to prove Theorem \ref{theorem: TW}. In Section \ref{const} we discuss
the multiplicative constant in the
asymptotics of $\det(I-K_{s}^{(k)})$ as $s\to-\infty$.

\section{Fredholm determinant in terms of a RH solution}\label{section fredholm}
In this section, we give a differential identity for the Fredholm determinant $\det(I-K_{s})$ in terms of the solution of a RH problem. This type of identity was obtained in a very general framework in \cite{IIKS, DIZ, BBIK, BD}. For the convenience of the reader, we recall briefly how to do it.

\medskip

Recall first that $K_s$ is an integral operator acting on $L^2(s,+\infty)$ with
kernel $K=K^{(k)}$ which we can write in the form
\begin{equation}\label{kernel fh}
K(u,v)=\frac{f^T(u)h(v)}{u-v}, \qquad f=\begin{pmatrix}\Phi_1\\
\Phi_2\end{pmatrix},\qquad h=\frac{1}{2\pi i}\begin{pmatrix}-\Phi_2\\
\Phi_1\end{pmatrix},
\end{equation}
where $\Phi_1$ and $\Phi_2$ are related to the Painlev\'e I hierarchy as explained in the introduction.
Note that
\begin{multline}\label{identityFredholmresolvent}
\frac{d}{ds}\ln\det(I-K_s)=-\tr\left((I-K_s)^{-1}{dK_s\over ds}\right)=
((I-K_s)^{-1}K_s)(s,s)\\
=((I-K_s)^{-1}(K_s-I+I))(s,s)=R_s(s,s),
\end{multline}
where $I+R_s$ is the resolvent of the operator $K_s$ given by
\begin{equation}
I+R_s=(I-K_s)^{-1}.
\end{equation}
The operator $R_s$ has a kernel of the form \cite[Lemma 2.8]{DIZ}:
\begin{equation}\label{resolvent}
R_s(u,v)=\frac{F^T(u)H(v)}{u-v}, \qquad F=(I-K_s)^{-1}f,\qquad H=(I-K_s)^{-1}h.
\end{equation}
The functions $F$ and $H$ can be expressed \cite{BBIK} in terms of the solution
\begin{equation}
Y(\zeta)=I-\int_s^{\infty} \frac{F(u)h^T(u)}{u-\zeta}du
\end{equation}
of the following RH problem.
\subsubsection*{RH problem for $Y$}

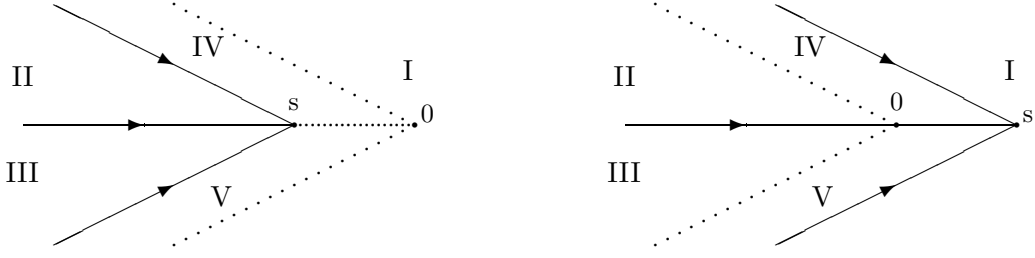
\begin{figure}[t]
\begin{center}
    \setlength{\unitlength}{0.8truemm}
    \begin{picture}(200,48.5)(0,2.5)
        \put(50,27.5){\thicklines\circle*{.8}}
        \put(49,30){\small s}
        \put(70,27.5){\thicklines\circle*{.8}}
        \put(71,28){\small 0}

        \put(50,27.5){\line(-2,1){40}}
        \multiput(70,27.5)(-2,1){21}{\circle*{0.1}}

        \put(50,27.5){\line(-2,-1){40}}
        \multiput(70,27.5)(-2,-1){21}{\circle*{0.1}}
        \put(5,27.5){\line(1,0){45}}
        \multiput(50,27.5)(1,0){20}{\circle*{0.1}}


        \put(68,35){${\rm I}$}
        \put(3,34){${\rm II}$}
        \put(2,18){${\rm III}$}

        \put(33,39){${\rm IV}$}
        \put(36,14){${\rm V}$}

        \put(30,37.5){\thicklines\vector(2,-1){.0001}}
        \put(30,17.5){\thicklines\vector(2,1){.0001}}
        \put(25,27.5){\thicklines\vector(1,0){.0001}}

        \put(150,27.5){\thicklines\circle*{.8}}
        \put(149,30){\small 0}
        \put(170,27.5){\thicklines\circle*{.8}}
        \put(171,28){\small s}

        \put(170,27.5){\line(-2,1){40}}
        \multiput(150,27.5)(-2,1){21}{\circle*{0.1}}

        \put(170,27.5){\line(-2,-1){40}}
        \multiput(150,27.5)(-2,-1){21}{\circle*{0.1}}
        \put(105,27.5){\line(1,0){65}}


        \put(168,35){${\rm I}$}
        \put(103,34){${\rm II}$}
        \put(102,18){${\rm III}$}

        \put(133,39){${\rm IV}$}
        \put(136,14){${\rm V}$}

        \put(150,37.5){\thicklines\vector(2,-1){.0001}}
        \put(150,17.5){\thicklines\vector(2,1){.0001}}
        \put(125,27.5){\thicklines\vector(1,0){.0001}}

    \end{picture}
    \caption{Regions I, II, III, and IV. In the case $s<0$, at the left, and in the case $s>0$, at the right.}
    \label{figure: Sigma1}
\end{center}
\end{figure}
\begin{itemize}
\item[(a)] $Y$ is analytic in $\mathbb C\setminus [s,+\infty)$.
\item[(b)] $Y(z)$ has $L^2$ boundary values related by the condition
$Y_+(x)=Y_-(x)v_Y(x)$ for $x\in(s,+\infty)$, with
\begin{equation}
v_Y(x)=I-2\pi i f(x)h^T(x).
\end{equation}
\item[(c)] $Y(\zeta)= I+\bigO(\zeta^{-1})$ as $\zeta\to \infty$.
\end{itemize}

\medskip

For $u\in\mathbb R$, the functions $F$ and $H$ given by (\ref{resolvent}) can be written as \cite[Lemma 2.12]{DIZ}
\begin{equation}\label{FH}
F(u)=Y_+(u)f(u), \qquad H(u)=(Y_+^{-1})^T(u)h(u),
\end{equation}
where $Y(u)$ is the solution of the above RH problem.

\medskip

The functions $\Phi_1^{(k)}$ and $\Phi_2^{(k)}$ now appear in the jump matrix $v_Y$.
A straightforward transformation produces a RH problem with constant jump matrices.
Namely, set (see Figure \ref{figure: Sigma1})
\begin{equation}
X(\zeta)=\begin{cases}\begin{array}{ll}Y(\zeta)\Phi(\zeta),&\mbox{ for $\zeta$ in region I, II, III,}\\
Y(\zeta)\Phi(\zeta)\begin{pmatrix}1&0\\1&1\end{pmatrix},&\mbox{ for $\zeta$ in region IV,}\\
Y(\zeta)\Phi(\zeta)\begin{pmatrix}1&0\\-1&1\end{pmatrix},&\mbox{ for $\zeta$ in region V,}
\end{array}
\end{cases}
\qquad\mbox{ if $s<0$,}
\end{equation}
and
\begin{equation}
X(\zeta)=\begin{cases}\begin{array}{ll}Y(\zeta)\Phi(\zeta),&\mbox{ for $\zeta$ in region I, II, III,}\\
Y(\zeta)\Phi(\zeta)\begin{pmatrix}1&0\\-1&1\end{pmatrix},&\mbox{ for $\zeta$ in region IV,}\\
Y(\zeta)\Phi(\zeta)\begin{pmatrix}1&0\\1&1\end{pmatrix},&\mbox{ for $\zeta$ in region V,}
\end{array}
\end{cases}
\qquad\mbox{ if $s>0$.}
\end{equation}

Then it is easy to verify that $X$ satisfies the following RH problem \cite{BBIK}.

\subsubsection*{RH problem for $X$}
\begin{itemize}
    \item[(a)] $X$ is analytic in $\mathbb{C}\setminus\Sigma^{(s)}$, where \begin{align*}&\Sigma^{(s)}=\Sigma^{(s)}_1\cup\Sigma^{(s)}_2\cup\Sigma^{(s)}_3\cup\{s\},\\
 &\Sigma^{(s)}_2=(-\infty,s), \quad \Sigma^{(s)}_1=s+e^{\frac{-i\pi}{4k+3}}\mathbb R^-, \quad \Sigma^{(s)}_3=s+e^{\frac{i\pi}{4k+3}}\mathbb R^-,\end{align*} oriented as in Figure \ref{figure: Sigma1}.
    \item[(b)]
The boundary values of $X$ are related by the jump conditions
    \begin{align}\label{RHP X:b1}
        X_+(\zeta)&=X_-(\zeta)
            \begin{pmatrix}
                1 & 0 \\
                1 & 1
            \end{pmatrix},&& \mbox{for $\zeta\in\Sigma^{(s)}_1\cup\Sigma^{(s)}_3$,}\\[1ex]
        \label{RHP X:b2}X_+(\zeta)&=X_-(\zeta)
            \begin{pmatrix}
                0 & 1 \\
                -1 & 0
            \end{pmatrix},&& \mbox{for
            $\zeta\in\Sigma^{(s)}_2$.}
    \end{align}
    \item[(c)] $X$ has the following asymptotic behavior as
    $\zeta\to\infty$:
    \begin{equation}\label{RHP X:c}
        X(\zeta)=\zeta^{-\frac{1}{4}\sigma_3}N\left(I+X_\infty\zeta^{-1/2}+\bigO(\zeta^{-1})\right)
        e^{-\theta(\zeta)\sigma_3},
    \end{equation}
    where $X_\infty$ is independent of $\zeta$ (its explicit form will not
be needed below),
 $N$ and $\theta$ are given by
    \begin{equation}\label{def: N}
        N=\frac{1}{\sqrt 2}
        \begin{pmatrix}
             1 & 1 \\
             -1 & 1
        \end{pmatrix}e^{-\frac{1}{4}\pi i\sigma_3},
    \end{equation}
    and
    \begin{equation}\label{def theta}
        \theta(\zeta;t_0,\ldots,t_{2k})=\frac{2}{4k+3}\zeta^{\frac{4k+3}{2}}-2\sum_{j=0}^{2k-1}\frac{
        (-1)^{j}t_{j}}{2j+1}\zeta^{\frac{2j+1}{2}}.
    \end{equation}
    \item[(d)] $X$ has the following behavior near $0$:
    \begin{equation}\label{RHP X:d}X(\zeta)=\bigO(\log|\zeta-s|),\qquad\mbox{ as $\zeta\to s$}.\end{equation}
\end{itemize}
To obtain (\ref{RHP X:c}) in the sectors IV and V, note that
\[
e^{-\th\si_3}\begin{pmatrix}
             1 & 0 \\
             \pm 1 & 1
        \end{pmatrix}
=\begin{pmatrix}
             1 & 0 \\
             \pm e^{2\th} & 1
        \end{pmatrix}
e^{-\th\si_3}
\]
and the real part of $\th(\zeta)$ in those sectors is negative for large $|\zeta|$.

%
%
%
%

\begin{remark}
The existence of a RH solution $X$ is, for real values of $s, t_0, \ldots, t_{2k-1}$, not an issue, since we constructed $X$ explicitly using $Y$ and $\Phi$. Concerning uniqueness of a solution $X$, we note that Remark \ref{remark uniqueness} also applies to the RH problem for $X$, so $X$ is only unique up to the left multiplication with a lower-triangular matrix of the form $\begin{pmatrix}1&0\\ \om&1\end{pmatrix}$.
\end{remark}
\begin{remark}\label{remark: first column}
Multiplying the three jump matrices for $X$ in counterclockwise direction, we obtain
\begin{equation}
\begin{pmatrix}
                1 & 0 \\
                1 & 1
            \end{pmatrix}\begin{pmatrix}
                0 & 1 \\
                -1 & 0
            \end{pmatrix}\begin{pmatrix}
                1 & 0 \\
                1 & 1
            \end{pmatrix}
            =\begin{pmatrix}
                1 & 1 \\
                0 & 1
            \end{pmatrix}.
\end{equation}
It follows from this observation that the first column of $X$
in the sector between $\Si_1^{(s)}$
and $(s,+\infty)$ has no branching. In other words, the elements in the
first column of $X$ can be extended from this sector
to entire functions in the complex plane.
\end{remark}

Now we will express the logarithmic derivative of $\det(I-K_s)$ in terms of $X$.
By (\ref{FH}), one verifies that
\[
F(\zeta)=\begin{pmatrix}X_1(\zeta)\\ X_2(\zeta)\end{pmatrix},\qquad
H(\zeta)=\frac{1}{2\pi i}\begin{pmatrix}-X_2(\zeta)\\ X_1(\zeta)\end{pmatrix},
\]
where $X_1$ and $X_2$ are the analytic extensions of the functions $X_{11}$ and $X_{21}$ from the sector between $\Si_1^{(s)}$
and $(s,+\infty)$ to the whole complex plane,
cf. Remark \ref{remark: first column}.
Using (\ref{resolvent}) with $v=s$ and $u\to s$, we then obtain the
following identity by (\ref{identityFredholmresolvent}):
\begin{equation}\label{identityFredholmX}
\frac{d}{ds}\ln\det(I-K_s)=\frac{1}{2\pi i}\left.\left(X^{-1}(\zeta)X'(\zeta)\right)_{21}\right|_{\zeta\to s},
\end{equation}
where the limit is taken as $\zeta$ approaches $s$ from the sector I if $s>0$
(sector IV if $s<0$), see Figure \ref{figure: Sigma1}.
This differential identity
is the starting point of our analysis, and is independent of the
chosen RH solution $X$. In the next section we will obtain
asymptotics for $X$ as $s\to -\infty$, which will lead to
asymptotics for the Fredholm determinant. In Section \ref{section
Painleve} we will find a Painlev\'e expression for the right hand side
of (\ref{identityFredholmX}).

\section{Large gap asymptotics}\label{section large gap}

We will now apply the steepest descent method
of Deift and Zhou
to the RH problem for $X$ when $s\to -\infty$. This approach is very similar to the one used by Kapaev in \cite{Kapaev}, and to techniques used several times in \cite{FIKN}.

\subsection{Rescaled RH problem}

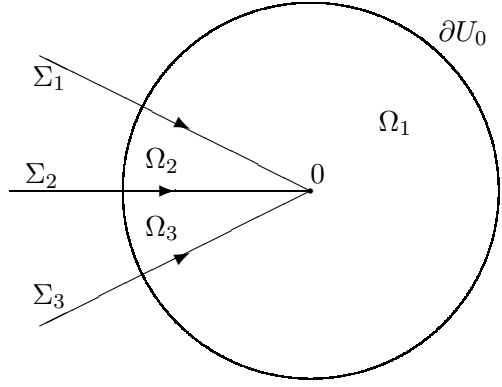
\begin{figure}[t]
\begin{center}
    \setlength{\unitlength}{1truemm}
    \begin{picture}(100,55)(-5,10)
        \cCircle(50,40){25}[f]
        \put(67,60){$\partial U_0$}
        \put(50,40){\thicklines\circle*{.8}}
        \put(50,41){$0$}
        \put(50,40){\line(-2,1){36}}
        \put(50,40){\line(-2,-1){36}}
        \put(50,40){\line(-1,0){40}}
        \put(34,48.2){\thicklines\vector(2,-1){.0001}}
        \put(34,31.8){\thicklines\vector(2,1){.0001}}
        \put(32,40){\thicklines\vector(1,0){.0001}}
        \put(13,54){$\Sigma_1$}
        \put(12,41){$\Sigma_2$}
        \put(13,25){$\Sigma_3$}
        \put(59,48){$\Omega_1$}
        \put(28,43){$\Omega_2$}
        \put(28,34){$\Omega_3$}
    \end{picture}
    \caption{Contour $\Sigma=\Sigma_1\cup\Sigma_2\cup\Sigma_3$, a circle $\partial U_0$ in which the local parametrix needs to be constructed, and the regions $\Omega_1, \Omega_2, \Omega_3$.}
    \label{newfig}
\end{center}
\end{figure}

Let us first consider the shifted and rescaled version of $X$:
\begin{equation}\label{def hatPhi}
T(\zeta;s,t_0,\ldots , t_{2k-1})=X(|s|\zeta+s;s,t_0,\ldots , t_{2k-1}).
\end{equation}
The shift with $s$ is convenient to 'center' the RH problem at the origin.
From the RH properties of $X$, we conclude that $T$ satisfies
the following RH problem (see Figure \ref{newfig}).
\subsubsection*{RH problem for $T$}
\begin{itemize}
    \item[(a)] $T$ is analytic in $\mathbb{C}\setminus\Sigma$, where
    \begin{align}
    &\label{def Sigma}\Sigma=\Sigma_1\cup\Sigma_2\cup\Sigma_3\cup\{0\}, \\
    &\Sigma_2=(-\infty,0), \quad \Sigma_1=e^{\frac{-i\pi}{4k+3}}\mathbb R^-, \quad \Sigma_3=e^{\frac{i\pi}{4k+3}}\mathbb R^-.\end{align}
    \item[(b)] $T$ has $L^2$ boundary values which
satisfy
    \begin{align}
        T_+(\zeta)&=T_-(\zeta)
            \begin{pmatrix}
                1 & 0 \\
                1 & 1
            \end{pmatrix},&& \mbox{for $\zeta\in\Sigma_1\cup\Sigma_3$,}\\[1ex]
        T_+(\zeta)&=T_-(\zeta)
            \begin{pmatrix}
                0 & 1 \\
                -1 & 0
            \end{pmatrix},&& \mbox{for
            $\zeta\in\Sigma_2$.}
    \end{align}
    \item[(c)] As
    $\zeta\to\infty$,
    \begin{equation}\label{RHP T:c1}
        T(\zeta)=(|s|\zeta)^{-\frac{1}{4}\sigma_3}N\left(I+|s|^{-1/2}X_\infty\zeta^{-1/2}+\bigO(\zeta^{-1})\right)
        e^{-\theta(|s|\zeta+s)\sigma_3},
    \end{equation}
    where $N$ and $\theta$ are given by (\ref{def: N}) and (\ref{def theta}), and
 $X_\infty$ is the matrix appearing in the expansion (\ref{RHP X:c})
for $X$ at infinity.
\end{itemize}


Expanding $\theta(|s|\zeta+s)$ as $\zeta\to\infty$ for fixed $s<0$, we obtain using (\ref{def theta}),
\begin{equation}\label{expansion theta}
\theta(|s|\zeta+s)=|s|^{\frac{4k+3}{2}}g(\zeta;s,t_0,\ldots, t_{2k-1})+
|s|^{\frac{4k+3}{2}}\gamma\zeta^{-1/2}+\bigO(\zeta^{-3/2}),
\end{equation}
where $g$ is equal to
\begin{equation}\label{def g}
g(\zeta;s,t_0,\ldots, t_{2k-1})=
\sum_{j=0}^{2k+1}(-1)^{j+1}a_j(s)\zeta^{j+\frac{1}{2}},
\end{equation}
with the principal branches for the fractional powers, and with the coefficients
$a_j$ given by
\begin{multline}
\label{def aj1}
a_j(s)={1\over\Gamma(j+\frac{3}{2})}
\left(
\frac{\Gamma(2k +\frac{3}{2})}{\Gamma(2k+2-j)}
+\sum_{m=2}^{2k+1-j}t_{2k+1-m}\frac{\Gamma(2k+\frac{3}{2}-m)}{\Gamma(2k+2-j-m)}
|s|^{-m}\right),\\
j=0,\dots,2k+1.
\end{multline}
(the sum vanishes for $j=2k$ and $j=2k+1$).
The explicit expression for
the coefficient $\gamma=\gamma(k,s,t_0,\dots,t_{2k-1})$
in (\ref{expansion theta}) will not be important below.

\subsection{Normalized RH problem}

The next step is to normalize the RH problem at infinity in a suitable way. We do this using the $g$-function constructed above. Set
\begin{equation}\label{ST}
S(\zeta)=T(\zeta)\exp\left\{|s|^{\frac{4k+3}{2}}g(\zeta)\sigma_3\right\},
\end{equation}
so that we have
\subsubsection*{RH problem for $S$}
\begin{itemize}
    \item[(a)] $S$ is analytic in $\mathbb{C}\setminus\Sigma$.
    \item[(b)] $S$ has $L^2$ boundary values that satisfy the jump
     relations $S_+=S_-v_S$ on
    $\Sigma$, with
    \begin{align}
        &v_S(\zeta)=
            \begin{pmatrix}
                1 & 0 \\
                \exp\left\{2|s|^{\frac{4k+3}{2}}g(\zeta)\right\} & 1
            \end{pmatrix},&& \mbox{for $\zeta\in\Sigma_1\cup\Sigma_3$,}\\[1ex]
        &v_S(\zeta)=
            \begin{pmatrix}
                0 & 1 \\
                -1 & 0
            \end{pmatrix},&& \mbox{for
            $\zeta\in\Sigma_2$.}
    \end{align}
    \item[(c)] $S$ has the following asymptotic behavior as
    $\zeta\to\infty$:
    \begin{equation}\label{RHP S:c1}
        S(\zeta)=(|s|\zeta)^{-\frac{1}{4}\sigma_3}N\left(I+S_\infty\zeta^{-1/2}+\bigO(\zeta^{-1})\right),
    \end{equation}
    with $N$ given by (\ref{def: N}), and \begin{equation}\label{S1}
    S_\infty(s)=|s|^{-1/2}X_\infty
        -\gamma|s|^{\frac{4k+3}{2}}\sigma_3,
    \end{equation}
    where $X_\infty$ is the matrix appearing in (\ref{RHP X:c}).
\end{itemize}

Using (\ref{def g})--(\ref{def aj1}),  one verifies that for sufficiently large
negative $s$
\begin{equation}\label{Reg}
\Re g(\zeta)<0,\qquad\mbox{ as $\zeta\in \Sigma_1\cup\Sigma_3$.}
\end{equation}
Note that the values of the coefficients $t_j$ do not matter for this simple but important observation, as long as $|s|$ is sufficiently large. The inequality (\ref{Reg}) implies that the jump matrix $v_S$ for $S$ is exponentially close to the identity matrix on $\Sigma_1\cup\Sigma_3$ as $s\to -\infty$.
For $\zeta$ near $0$, this uniform convergence breaks down.

\subsection{Outside parametrix}

Ignoring the exponentially small jumps and a neighborhood of $0$, we are led to a RH problem, which we refer to as the RH problem for the outside parametrix
$P^{(\infty)}$:

\subsubsection*{RH problem for $P^{(\infty)}$}
\begin{itemize}
    \item[(a)] $P^{(\infty)}$ is analytic in $\mathbb{C}\setminus (-\infty,0]$.
    \item[(b)] $P^{(\infty)}$ has $L^2$ boundary values that satisfy
    the following relation on
    $(-\infty,0)$:
    \begin{align}
        P^{(\infty)}_+(\zeta)&=P^{(\infty)}_-(\zeta)
            \begin{pmatrix}
                0 & 1 \\
                -1 & 0
            \end{pmatrix}.
    \end{align}
    \item[(c)] As
    $\zeta\to\infty$,
    \begin{equation}\label{RHP Pinfty:c}
        P^{(\infty)}(\zeta)=(|s|\zeta)^{-\frac{1}{4}\sigma_3}N\left(I+\bigO(\zeta^{-1/2})\right).
    \end{equation}
\end{itemize}
A solution $P^{(\infty)}$ is given by simply removing the error terms in
(\ref{RHP Pinfty:c}):
\begin{equation}\label{Pin}
P^{(\infty)}(\zeta)=(|s|\zeta)^{-\frac{1}{4}\sigma_3}N.
\end{equation}

\subsection{Local parametrix}
Consider a sufficiently small disk $U_0$ (of fixed radius, independent of $s$)
centered at the origin.
The goal of this section is to construct a function $P$ in $U_0$ which satisfies the same jump conditions as $S$, and matches with the outside parametrix $P^{(\infty)}$
for large negative $s$ at the boundary $\partial U_0$ of $U_0$. Thus
we have
\subsubsection*{RH problem for $P$}\label{subsubsectionP}
\begin{itemize}
    \item[(a)] $P$ is analytic in $\overline{U_0}\setminus\Sigma$.
    \item[(b)] $P$ has $L^2$ boundary values that satisfy the following
    jump relations on $U_0\cap \Sigma$:
    \begin{align}\label{RHP P:b1}
        P_+(\zeta)&=P_-(\zeta)
            \begin{pmatrix}
                1 & 0 \\
                \exp\left\{2|s|^{\frac{4k+3}{2}}g(\zeta)\right\} & 1
            \end{pmatrix},&& \mbox{for $\zeta\in U_0\cap\left(\Sigma_1\cup\Sigma_3\right)$,}\\[1ex]
            \label{RHP P:b2}
        P_+(\zeta)&=P_-(\zeta)
            \begin{pmatrix}
                0 & 1 \\
                -1 & 0
            \end{pmatrix},&& \mbox{for
            $\zeta\in U_0\cap \Sigma_2$.}
    \end{align}
    \item[(c)] As $s\to -\infty$, $P$ satisfies the following matching condition
     with $P^{(\infty)}$ at the boundary $\partial U_0$ of $U_0$:
    \begin{equation}\label{RHP P:c1}
        P(\zeta)=P^{(\infty)}(\zeta)\left(I+o(1)\right),\qquad\mbox{ for $\zeta\in\partial U_0$.}
    \end{equation}
\end{itemize}

This local parametrix will approximate $S$ near the origin, while the outside parametrix approximates $S$ away from the origin. We now construct the local parametrix $P$ explicitly
using Bessel and Hankel functions.

\subsubsection{Bessel model RH problem}
Inspired by the constructions in e.g.\ \cite{KV1, DIK}, we
define the functions $J_1$, $J_2$, and $J_3$:
\begin{align}
&\label{def J1}J_1(\lambda)=e^{-\frac{\pi i}{4}\sigma_3}\pi^{\frac{1}{2}\sigma_3}{\small\begin{pmatrix}I_0(\lambda^{1/2})&\frac{i}{\pi}K_0(\lambda^{1/2})\\ \pi i\lambda^{1/2}I_0'(\lambda^{1/2})&-\lambda^{1/2}K_0'(\lambda^{1/2})\end{pmatrix}},\\
&\label{def J2}J_2(\lambda)=\frac{1}{2}e^{-\frac{\pi i}{4}\sigma_3}\pi^{\frac{1}{2}\sigma_3}{\small\begin{pmatrix}H_0^{(1)}(-i\lambda^{1/2})&H_0^{(2)}(-i\lambda^{1/2})\\ \pi \lambda^{1/2}H_0^{(1)'}(-i\lambda^{1/2})&\pi\lambda^{1/2}H_0^{(2)'}(-i\lambda^{1/2})\end{pmatrix}},\\
&\label{def J3}J_3(\lambda)=\frac{1}{2}e^{-\frac{\pi i}{4}\sigma_3}\pi^{\frac{1}{2}\sigma_3}{\small\begin{pmatrix}H_0^{(2)}(i\lambda^{1/2})&-H_0^{(1)}(i\lambda^{1/2})\\ -\pi \lambda^{1/2}H_0^{(2)'}(i\lambda^{1/2})&\pi\lambda^{1/2}H_0^{(1)'}(i\lambda^{1/2})\end{pmatrix}},
\end{align}
where $I_0$, $K_0$, and $H_0^{(j)}$ denote the usual modified Bessel functions and Hankel functions \cite{AS} defined on a plane (rather than universal covering)
with the cut along $(-\infty,0]$, and where we take principal branches of $\lambda^{1/2}$, analytic off $(-\infty,0]$ and positive for $\lambda>0$.
The functions $J_1$, $J_2$, and $J_3$ are related to each other by the constant matrices:
\begin{align}
            \label{RHP J: b1}
            J_1(\lambda) &= J_2(\lambda)
                \begin{pmatrix}
                    1 & 0\\
                    1 & 1
                \end{pmatrix},&&\mbox{ for $\lambda\in\mathbb C\setminus(-\infty,0]$,}\\[1ex]
            \label{RHP J: b2}
            J_2(\lambda) &= J_3(\lambda)
                \begin{pmatrix}
                    0 & 1\\
                    -1 & 0
                \end{pmatrix},&&\mbox{ for $\lambda<0$,}
                \\[1ex]
            \label{RHP J: b3}
            J_3(\lambda) &= J_1(\lambda)
                \begin{pmatrix}
                    1 & 0\\
                    1 & 1
                \end{pmatrix},&&\mbox{ for $\lambda\in\mathbb C\setminus(-\infty,0]$,}
        \end{align}
and have the following large $\lambda$ asymptotics:
\begin{multline}\label{RHP J: c}
            J_m(\lambda)=\lambda^{-\frac{1}{4}\sigma_3}N{\small\left[I+\frac{1}{8\sqrt\lambda}
            \begin{pmatrix}-1&-2i\\-2i&1\end{pmatrix}+
         \bigO(\lambda^{-1})\right]}e^{\lambda^{1/2}\sigma_3},\\ \qquad
                \mbox{uniformly as $\lambda\to\infty$ in sector $S_m$,}
        \end{multline}
where \begin{align}\label{sectorsJ}
&S_1=\{\lambda: -\pi +\epsilon <\arg\lambda< \pi -\epsilon\},\\
&S_2=\{\lambda: -\pi +\epsilon <\arg\lambda< \pi\},\\
&S_3=\{\lambda: -\pi <\arg\lambda< \pi -\epsilon\}.
\end{align}
Using these functions, we can now construct the local parametrix.

\subsubsection{Construction of the local parametrix}

We define the parametrix $P$ in the form
\begin{equation}\label{P}
P(\zeta)=E(\zeta)J_m(|s|^{4k+3}f(\zeta))\exp\left\{
|s|^{\frac{4k+3}{2}}g(\zeta)\sigma_3\right\}, \qquad\mbox{ for $\zeta\in\Omega_m$,}
\end{equation}
where $E(\zeta)$ is an analytic function to be determined later on,
\begin{align*}
&\Omega_1=\{-\frac{4k+2}{4k+3}\pi<\arg\zeta<\frac{4k+2}{4k+3}\pi\}\cap U_0, \\
&\Omega_2=\{\frac{4k+2}{4k+3}\pi<\arg\zeta<\pi\}\cap U_0,\\
&\Omega_3=\{-\pi<\arg\zeta<-\frac{4k+2}{4k+3}\pi\}\cap U_0
\end{align*}
(see Figure \ref{newfig}),
and $f$ is defined as follows:
\begin{equation}
f(\zeta)=g(\zeta)^2=\zeta \left(a_0+\sum_{j=1}^{2k+1}(-1)^{j}a_j(s)\zeta^{j}\right)^2,
\end{equation}
so that $f$ is a polynomial in $\zeta$ with
\begin{equation}
f(0)=0,\qquad f'(0)=a_0^2(s)>0.
\end{equation}
Choosing $U_0$ sufficiently small,
we see that $f(\zeta)$ is a conformal mapping of $U_0$ which maps zero to zero,
preserves the real line, and maps $\Omega_j$ into $S_j$ for $j=1,2,3$, respectively.
Thus we can use the asymptotic expansion (\ref{RHP J: c}) and the jump relations (\ref{RHP J: b1})--(\ref{RHP J: b3}) for $J_m(|s|^{4k+3}f(\zeta))$.
Namely, using (\ref{RHP J: b1})--(\ref{RHP J: b3}) and the fact that $g_+=-g_-$ on $(-\infty,0)$, we deduce from (\ref{P}) that $P$ satisfies
the jump conditions
(\ref{RHP P:b1}), (\ref{RHP P:b2}) on $\Sigma\cap U_0$.

Finally, define the analytic pre-factor $E$ by
\begin{equation}\label{def E}
E(\zeta)=|s|^{(k+\frac{1}{2})\si_3}
\zeta^{-\frac{1}{4}\sigma_3}f(\zeta)^{\frac{1}{4}\sigma_3}=
|s|^{(k+\frac{1}{2})\si_3}
 \left(a_0+\sum_{j=1}^{2k+1}(-1)^{j}a_j(s)\zeta^{j}\right)^{\si_3/2}.
\end{equation}

From (\ref{def aj1}) we see that $f(\zeta)=f(\zeta,s)$ tends to some $f_0(\zeta)$ independent of $s$
as $|s|\to\infty$.
Therefore,
for $\zeta\in\partial U_0$, we can use the asymptotic expansion
(\ref{RHP J: c}) as $s\to -\infty$ to conclude from (\ref{def E}) that
\begin{multline}\label{matching P Pinfty}
P(\zeta)=P^{(\infty)}(\zeta)\left[I+\frac{1}{8|s|^{\frac{4k+3}{2}}f(\zeta)^{1/2}}
            \begin{pmatrix}-1&-2i\\-2i&1\end{pmatrix}+
         \bigO(|s|^{-(4k+3)})\right].
\end{multline}
It follows that $P$ indeed satisfies the RH conditions of the form proposed at the beginning of Section \ref{subsubsectionP}.

\subsection{Final transformation}

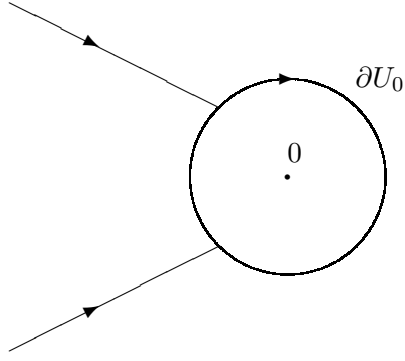
\begin{figure}[t]
\begin{center}
    \setlength{\unitlength}{1truemm}
    \begin{picture}(100,55)(-5,10)
        \cCircle(50,40){13}[f]
        \put(50,40){\thicklines\circle*{.8}}
        \put(50,42){$0$}
        \put(59,52){$\partial U_0$}
        \put(41,49.2){\line(-2,1){28}}
        \put(41,30.8){\line(-2,-1){28}}
        \put(25,57.2){\thicklines\vector(2,-1){.0001}}
        \put(25,22.8){\thicklines\vector(2,1){.0001}}
        \put(51,53){\thicklines\vector(1,0){.0001}}
    \end{picture}
    \caption{Reduced system of contours $\hat\Sigma_R$ independent of $s$.}
    \label{figure: SigmaR}
\end{center}
\end{figure}

Let us define
\begin{align}\label{def R1}
&R(\zeta)=\begin{pmatrix}1&0\\
-|s|^{1/2}(NS_\infty N^{-1})_{21}&1\end{pmatrix}S(\zeta)P(\zeta)^{-1},&&\mbox{for $\zeta\in U_0$,}\\
&\label{def R2}R(\zeta)=\begin{pmatrix}1&0\\
-|s|^{1/2}(NS_\infty N^{-1})_{21}&1\end{pmatrix}S(\zeta)P^{(\infty)}(\zeta)^{-1},&&\mbox{for $\zeta\in \mathbb C\setminus U_0$,}
\end{align}
where $S_\infty$ is the matrix appearing in the asymptotic expansion
for $S$ and given by (\ref{S1}), and $N$ is given by (\ref{def: N}).
Then we have

\subsubsection*{RH problem for $R$}\label{section: large gap final}
\begin{itemize}
\item[(a)] $R$ is analytic in $\mathbb C\setminus \Sigma_R$, with $\Sigma_R$ as shown in Figure
\ref{figure: SigmaR}.
\item[(b)] $R$ has $L^2$ boundary values satisfying
the jump relations $R_+=R_-v_R$ on $\Sigma_R$, with
\begin{align}
&v_R(\zeta)=P(\zeta)P^{(\infty)}(\zeta)^{-1},&&\mbox{ for $\zeta\in \partial U_0$,}\\
&v_R(\zeta)=P^{(\infty)}(\zeta)v_S(\zeta)P^{(\infty)}(\zeta)^{-1},
&&\mbox{ for $\zeta\in\Sigma_R\setminus\overline{U_0}$.}
\end{align}
\item[(c)] $R(\zeta)\to I$ as $\zeta\to\infty$.
\end{itemize}
Outside $\overline U_0$, we have for some $c>0$,
\[v_R(\zeta)=
P^{(\infty)}(\zeta)
v_S(\zeta)P^{(\infty)}(\zeta)^{-1}
=I+\bigO(e^{-c|s|^{\frac{4k+3}{2}}}),\qquad\mbox{ as $s\to -\infty$.}\]
For $\zeta\in\partial U_0$, it follows from (\ref{matching P Pinfty}) that in the limit as $s\to -\infty$ we have
\begin{equation}\label{expansion JR}
v_R(\zeta)=I+|s|^{-\frac{1}{4}\sigma_3}v_1(\zeta)|s|^{\frac{1}{4}\sigma_3}|s|^{-\frac{4k+3}{2}}+\bigO(|s|^{-4k-5/2}),\end{equation}
where, by (\ref{matching P Pinfty}),
\begin{align}
&v_1(\zeta)=\frac{1}{8}\begin{pmatrix}0&-\frac{1}{\zeta^{1/2}f(\zeta)^{1/2}}\\
3\frac{\zeta^{1/2}}{f(\zeta)^{1/2}}&0\end{pmatrix}.
\end{align}
By standard analysis, we see that $R$ also has an asymptotic expansion
of the form
\begin{equation}\label{expansion R}
R(\zeta)=I+|s|^{-\frac{1}{4}\sigma_3}R_1(\zeta)|s|^{\frac{1}{4}\sigma_3}|s|^{-\frac{4k+3}{2}}+\bigO(|s|^{-4k-5/2}).
\end{equation}
This expansion is uniform in $\zeta$, and the l.h.s. is analytic,
 so that one can differentiate these asymptotics with respect to $\zeta$.

Collecting coefficients at the same powers of $|s|$ in the jump relation $R_+=R_-v_R$,
we obtain:
\begin{align}
&\label{jump R1}R_{1,+}(\zeta)=R_{1,-}(\zeta)+v_1(\zeta),&&\mbox{ for $\zeta\in\partial U_0$,}
\end{align}
where $R_1$ is analytic in $\mathbb C\setminus \partial U_0$. Since $R\to I$
as $|s|\to\infty$, it follows that $R_1$ tends to $0$ at infinity. Together with the jump condition (\ref{jump R1}) this normalization enables us to compute $R_1$ explicitly:
\begin{equation}
R_1(\zeta)={1\over 2\pi i}\int_{\partial U_0}\frac{v_1(z)}{z-\zeta}dz=
\begin{cases}
-\mathcal H(v_1(\zeta)),&\mbox{for $\zeta\in U_0$,}\cr
\frac{1}{\zeta}\Res (v_1(\zeta);0),&\mbox{for $\zeta\in\mathbb C\setminus U_0$,}
\end{cases}
\end{equation}
where $\mathcal H(f)$ denotes the Taylor part of the Laurent expansion of $f$
near $0$. Here we used the fact that the pole of $v_1(\zeta)$ is simple.

It is straightforward to check that
\begin{align}
&\label{R10}R_1(0)=\begin{pmatrix}0&\frac{a_1}{8a_0^2}\\ -\frac{3}{8a_0}&0\end{pmatrix},\\
&\label{R1p}R_1'(0)=\begin{pmatrix}0&*\\
-\frac{3a_1}{8a_0^2}&0
\end{pmatrix}.
\end{align}

\subsection{Large gap asymptotics}

Using (\ref{identityFredholmX}) and the definition (\ref{def hatPhi}) of $T$, we obtain the following identity:
\begin{equation*}
F(s):=\frac{d}{ds}\ln\det(I-K_s^{(k)})=\frac{1}{2\pi i|s|}
\left.\left(T^{-1}T_\zeta'\right)_{21}\right|_{\zeta\to 0},
\end{equation*}
where $\zeta\to 0$ in Sector IV,
which by (\ref{ST}) leads to a similar identity in terms of the normalized
RH solution $S$,
\begin{equation*}
F(s)=\frac{1}{2\pi i|s|}\left.\left(S^{-1}S_\zeta'\right)_{21}\right|_{\zeta\to 0},
\end{equation*}
and by (\ref{def R1}),
\begin{equation*}
F(s)=\frac{1}{2\pi i|s|}\left.\left(P^{-1}P_\zeta'\right)_{21}\right|_{\zeta\to 0}+
\frac{1}{2\pi i|s|}\left.\left(P^{-1}R^{-1}R_\zeta' P\right)_{21}\right|_{\zeta\to 0}.
\end{equation*}
Substituting the parametrix given by (\ref{P}) into this expression, we find
\begin{multline}\label{F-threeterms}
F(s)=\frac{a_0^2(s)|s|^{4k+2}}{2\pi i}\left.\left(J^{-1}J_\lambda'\right)_{21}\right|_{\lambda\to 0}+\frac{1}{2\pi i|s|}\left.\left(J^{-1}E^{-1}E_\zeta'J\right)_{21}\right|_{\zeta\to 0}\\ +\frac{1}{2\pi i|s|}\left.\left(J^{-1}E^{-1}R^{-1}R_\zeta'EJ\right)_{21}\right|_{\zeta\to 0}.
\end{multline}
To analyze the first term, we use (\ref{def J1}) and the expansions of Bessel functions near the origin (see e.g.\ \cite{AS}) to conclude that
\begin{equation}\label{first term}
\frac{a_0^2(s)|s|^{4k+2}}{2\pi i}\left.\left(J^{-1}J_\lambda'\right)_{21}\right|_{\lambda\to 0}
=\frac{1}{4}a_0^2(s)|s|^{4k+2}.
\end{equation}
This is the leading order term in the asymptotic expansion for $\frac{d}{ds}\ln\det(I-K_s)$.
Next we obtain using (\ref{def E}) that
\begin{equation}\label{second term}
\frac{1}{2\pi i|s|}\left.\left(J^{-1}E^{-1}E_\zeta'J\right)_{21}\right|_{\zeta\to 0}=0.
\end{equation}
Computing the last term in (\ref{F-threeterms}) is slightly more involved, as we need to use the asymptotic expansion for $R$ here. We have
\begin{multline*}\label{third term}
\frac{i}{2\pi |s|}\left.\left(J^{-1}E^{-1}R^{-1}R_\zeta' EJ\right)_{21}\right|_{\zeta\to 0}
\\
\left.
=\frac{i}{2\pi |s|}\left(J^{-1}\left(|s|^{2k+1}a_0(s)\right)^{-\sigma_3/2}R(0)^{-1}R_\zeta'(0)
\left(|s|^{2k+1}a_0(s)\right)^{\sigma_3/2} J\right)_{21}\right|_{\ze\to 0}.
\end{multline*}
Let us now take a closer look at $R(0)^{-1}R_\zeta'(0)$. Using the asymptotic expansion
(\ref{expansion R}) for $R$ we see that
\begin{equation*}
R(0)^{-1}R_\zeta'(0)=
|s|^{-\sigma_3/4}R_1'(0)|s|^{\sigma_3/4}|s|^{-\frac{4k+3}{2}}+\bigO(|s|^{-4k-\frac{5}{2}}),
\end{equation*}
from which it follows by (\ref{R1p}) and the Bessel expansions near $0$ that
\begin{equation}\label{third term2}
\frac{1}{2\pi i|s|}\left.\left(J^{-1}E^{-1}R^{-1}R_\zeta' EJ\right)_{21}\right|_{\zeta\to 0}=\frac{3a_1(s)}{16a_0(s)|s|}+\bigO(|s|^{-\frac{4k+5}{2}}).
\end{equation}
Summing up (\ref{first term}), (\ref{second term}), and (\ref{third term2}), we obtain
\begin{equation}
\frac{d}{ds}\ln F(s)=\frac{1}{4}a_0^2(s)|s|^{4k+2}+\frac{3a_1(s)}{16a_0(s)|s|}+\bigO(|s|^{-\frac{4k+5}{2}}),
\end{equation}
which proves Theorem \ref{theorem: large gap}.

\section{A solution to the Painlev\'e II hierarchy}\label{section q}

In this section, we study the RH problem associated with the $\P2n$ equation (\ref{PIIn}) for $n$ odd. First we pose the RH problem in full generality, with a contour consisting of $4n+2$ rays connecting $0$ with infinity. Afterwards we consider one particular $\P2n$-solution $q$ with $\alpha=\frac{1}{2}$, for which the jump contour reduces to four rays.
We will then argue why $q$ has no poles for real values of $(x;\tau_1, \ldots, \tau_{n-1})$, and why it is real.
Afterwards we will perform a rather standard steepest descent analysis of the RH problem, which is similar to the one used in Section \ref{section large gap}, in order to find asymptotics for $q$ at $+\infty$. We will also obtain asymptotics for $q(x)$ as $x\to -\infty$. This part requires the construction of a $g$-function and is less straightforward.

\subsection{Lax system for the Painlev\'e II hierarchy}

The $\P2n$ equation (\ref{PIIn}) arises as the underlying compatibility condition of a linear system. This Lax system was found by Flaschka and Newell in \cite{FN} for the Painlev\'e II equation, and generalized to the Painlev\'e II hierarchy in \cite{CJM, Kudryashov, MazzoccoMo}.

\medskip

The linear system consists of the following equations for $2\times 2$ matrices $Z=Z(\ze;x,\tau_1, \ldots, \tau_{n-1})$:
\begin{equation}\label{Lax PII}
Z'_\ze=MZ, \qquad Z'_x=B_0Z,\qquad Z'_{\tau_j}
=B_jZ,\quad j=1,\ldots,n-1,
\end{equation}
where
the logarithmic derivatives with respect to $\ze$ and $x$ have the form
\begin{align}
&\label{M}M=\sum_{j=0}^{2n}M_{j}\ze^j -x\sigma_3+\frac{\alpha}{\ze}\sigma_1,\\
&\label{B}B_0=\begin{pmatrix}
-\ze&q\\q&\ze
\end{pmatrix},
\end{align}
and the $B_j$'s are polynomials of degree $2j+1$ in $\ze$.
The coefficients $M_j$ are independent of $\ze$. Except for (\ref{B}) above,
the explicit expressions for $M_j$ and $B_j$ in terms of $\ze$, $q$, and
$\tau_j$ will not be needed below.

The system of Lax equations (\ref{Lax PII}) is only compatible if $q$ satisfies the ${\rm P}_{{\rm II}}^{(n)}$ equation. Equivalently, in order to preserve the monodromy at infinity for the system $Z_\ze'=MZ$ when varying $x, \tau_1, \ldots , \tau_{n-1}$, it is necessary that $q$ solves the ${\rm P}_{{\rm II}}^{(n)}$ equation.
Using canonical solutions to the Lax system for a given $\P2n$-solution $q$, one can build a RH problem. We do this in the next subsection.

\subsection{The RH problem for the Painlev\'e II hierarchy}

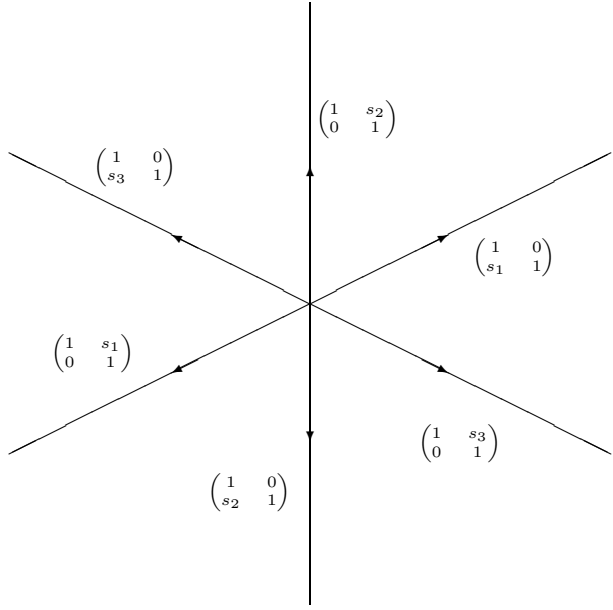
\begin{figure}[t]
\begin{center}
\setlength{\unitlength}{0.8truemm}
\begin{picture}(100,100)(0,0)
\put(77,57){{\tiny $\begin{pmatrix}1&0\\
s_1&1\end{pmatrix}$}}
\put(51,80){{\tiny $\begin{pmatrix}1&s_2\\
0&1\end{pmatrix}$}}
\put(14,72){{\tiny $\begin{pmatrix}1&0\\
s_3&1\end{pmatrix}$}}
\put(7,41){{\tiny $\begin{pmatrix}1&s_1\\
0&1\end{pmatrix}$}}
\put(33,18){{\tiny $\begin{pmatrix}1&0\\
s_2&1\end{pmatrix}$}}
\put(68,26){{\tiny $\begin{pmatrix}1&s_3\\
0&1\end{pmatrix}$}}
\put(50,50){\vector(0,1){23}} \put(50,50){\vector(0,-1){23}}
\put(50,73){\line(0,1){27}} \put(50,27){\line(0,-1){27}}
\put(50,50){\vector(2,1){23}} \put(50,50){\vector(2,-1){23}}
\put(73,61.5){\line(2,1){27}} \put(73,38.5){\line(2,-1){27}}
\put(50,50){\vector(-2,1){23}} \put(50,50){\vector(-2,-1){23}}
\put(27,61.5){\line(-2,1){27}} \put(27,38.5){\line(-2,-1){27}}
\end{picture}
\end{center}
\caption{Jumps for $\Psi(\zeta)$ if $n=1$. \label{fig:jumpsPsi}}
\end{figure}

For $j=1, \ldots, 4n+2$, let $\Omega_j$ be the sector
\begin{equation} \label{sector}
\Omega_j=\left\{\zeta\in\mathbb C \mid
    \frac{2j-3}{4n+2} \pi < \arg \zeta <
    \frac{2j-1}{4n+2} \pi \right\},
\end{equation}
and denote the boundaries of these sectors
\begin{equation} \label{Gammaj}
\Upsilon_j = \left\{ \zeta \mid \arg \zeta = \frac{2j-1}{4n+2} \pi \right\}
\quad \textrm{ for }j=1, \ldots, 4n+2,
\end{equation}
each of them oriented away from the origin. Set
$\Upsilon=\cup_{j=1}^{4n+2}\Upsilon_j$.
As described in \cite{FN, FIKN}, for a given solution $q$ to $\P2n$ one can choose $4n+2$ canonical solutions $Z_j$, $j=1,\dots,4n+2$ to the Lax system (\ref{Lax PII}) which have the asymptotic behavior
\[Z_j(\zeta) =\left(I+O(\zeta^{-1})\right)e^{-i\Theta(-i\zeta)\sigma_3},\qquad\mbox{ as $\zeta\to\infty$ in the sector $\Omega_j$},\]
with
\[\Theta(\zeta;x,\tau_1, \ldots,\tau_{n-1})=
\frac{(-1)^{n+1}}{4n+2}(2\zeta)^{2n+1}+
\sum_{j=1}^{n-1}\frac{(-1)^{j+1}
        \tau_{j}}{4j+2}(2\zeta)^{2j+1}+x\zeta.
 \]

From now on for the rest of the paper we assume that $n$ is odd.
Then the function defined as
\begin{equation}
\Psi(\zeta)=\Psi_j(\zeta)=Z_j(i\zeta),
\qquad\mbox{ for $\zeta\in\Omega_j$,}\qquad j=1,\dots,4n+2
\end{equation}
satisfies the following RH conditions.

\begin{itemize}
\item[(a)] $\Psi(\zeta)$ is analytic in
$\mathbb C \setminus \Upsilon$.
\item[(b)] $\Psi(\zeta)$ has continuous boundary values satisfying
$\Psi_+ (\zeta)= \Psi_- (\zeta) I_j$ for $\zeta\in\Upsilon_j\setminus \{ 0\}$
with
\begin{align}\label{Stokesmatrices}
&I_{2j+1}=\begin{pmatrix}1&0\\s_{2j+1}&1\end{pmatrix}, &\mbox{ for $j=0, \ldots , n$},\\
&I_{2j}=\begin{pmatrix}1&s_{2j}\\0&1\end{pmatrix}, &\mbox{ for $j=1, \ldots , n$},\\
&I_{2n+1+j}=I_{j}^T, &\mbox{ for $j=1, \ldots , 2n+1$},
\end{align}
\item[(c)] As $\zeta\to\infty$, we have
\begin{equation}\label{RHP Psi: c}\Psi(\zeta) =\left(I+\frac{\Psi_\infty}{\zeta}+O(\zeta^{-2})\right)e^{-i\Theta(\zeta)\sigma_3},\end{equation} with
\begin{equation}\label{def Theta}
        \Theta(\zeta;x,\tau_1, \ldots,\tau_{n-1})=\frac{1}{4n+2}(2\zeta)^{2n+1}+
\sum_{j=1}^{n-1}\frac{(-1)^{j+1}
        \tau_{j}}{4j+2}(2\zeta)^{2j+1}+x\zeta.
    \end{equation}
\end{itemize}

\begin{remark}\label{remark freedom contour}
The asymptotic behavior of the functions $\Psi_j$ is valid in sectors which are wider than $\Omega_j$. The precise shape of the contour $\Upsilon$ is thus not crucial. It is important that $\Upsilon_j$ is a curve connecting $0$ with infinity which lies asymptotically in the Stokes sector $\{\zeta: \frac{2j-2}{4n+2}\pi<\arg\zeta < \frac{2j}{4n+2}\pi\}$. For simplicity, we have chosen $\Upsilon$ so that it coincides with the anti-Stokes lines on which the leading order term of $\Theta(\zeta)$ (as $\zeta\to\infty$) is purely imaginary.
\end{remark}

Although we suppress this in our notation, $\Psi$ depends not only on $\zeta$ but also on the parameters $n$, $\alpha$, $x$, $\tau_1, \ldots,\tau_{2k}$ (which are present in the $\P2n$ equation (\ref{PIIn})) and on the Stokes multipliers $(s_1, \ldots, s_{2n+1})$.
If, for any set of Stokes multipliers $(s_1, \ldots, s_{2n+1})$, the above RH problem has a solution, it follows from the Lax system for $\Psi$ that
\begin{equation}\label{qPsi}
q(x;\tau_1, \ldots , \tau_{2k}):=2i\Psi_{\infty,12}(x,\tau_1, \ldots , \tau_{2k})=-2i\Psi_{\infty,21}(x,\tau_1, \ldots , \tau_{2k})
\end{equation}
solves the ${\rm P}_{{\rm II}}^{(n)}$ equation (\ref{PIIn}).

\medskip

However, the RH problem can only be solved if the Stokes multipliers satisfy
the condition
\begin{equation}\label{cyclic condition}
\Tr(S_1S_2\ldots S_{2n+1}\sigma_1)=-2i\sin\pi\alpha.
\end{equation}
For any set of Stokes multipliers lying on the ``monodromy surface''
given by (\ref{cyclic condition}), the RH problem has a unique solution which is meromorphic in $x$ and $\tau_1, \ldots , \tau_{2k}$; it determines
$q(x)$ by (\ref{qPsi}). The (isolated) values of
$(x,\tau_1, \ldots , \tau_{2k})$ for which the RH problem is not solvable
correspond to the poles of the $\P2n$-solution.
This map between the sets of Stokes multipliers $(s_1, \ldots, s_{2n+1})$
lying on the monodromy surface and the
solutions of the $\P2n$ equation (\ref{PIIn}) is one-to-one.

\subsection{Special solution to the Painlev\'e II hierarchy}\label{section: RHP}
The solution of $\P2n$ which is of interest to us, corresponds to the value of $\alpha=\frac{1}{2}$ and the set of Stokes multipliers
\begin{align}\label{Stokesmultipliers}
&s_1=s_{2n+1}=-i,\\
&s_2=s_3= \cdots =s_{2n}=0.
\end{align}
Because of the trivial jumps on the rays $\Upsilon_2, \ldots , \Upsilon_{2n}$, our jump contour is now reduced to
\[\Upsilon=\Upsilon_1\cup\Upsilon_{2n+1}\cup\Upsilon_{2n+2}\cup\Upsilon_{4n+2}.\]
Consider the following RH problem.

\subsubsection*{RH problem for $\Psi$:}
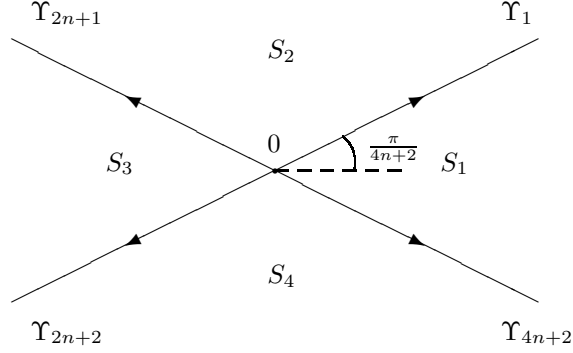
\begin{figure}[t]
\begin{center}
    \setlength{\unitlength}{1truemm}
    \begin{picture}(100,48.5)(0,2.5)
        \put(72,27.5){\small $S_1$}
        \put(49,42.5){\small $S_2$}
        \put(27.5,27.5){\small $S_3$}
        \put(49,12.5){\small $S_4$}

        \put(80,47.5){\small $\Upsilon_1$}
        \put(17.5,47.5){\small $\Upsilon_{2n+1}$}
        \put(17.5,5){\small $\Upsilon_{2n+2}$}
        \put(80,5){\small $\Upsilon_{4n+2}$}

        \put(50,27.5){\thicklines\circle*{.8}}
        \put(49,30){\small 0}
        \multiput(50,27.5)(3,0){6}{\line(1,0){1.75}}
        \qbezier(60.5,27.5)(61,30.5)(59,32)
        \put(62.25,30){\scriptsize $\frac{\pi}{4n+2}$}

        \put(15,10){\line(2,1){70}}
        \put(15,45){\line(2,-1){70}}
        \put(70,37.5){\thicklines\vector(2,1){.0001}}
        \put(70,17.5){\thicklines\vector(2,-1){.0001}}
        \put(30,37.5){\thicklines\vector(-2,1){.0001}}
        \put(30,17.5){\thicklines\vector(-2,-1){.0001}}
    \end{picture}
    \caption{Contour $\Upsilon$}
    \label{figure: RHP Psi}
\end{center}
\end{figure}
\begin{itemize}
\item[(a)] $\Psi(\zeta)$ is analytic in $\mathbb C\setminus\Upsilon$.
\item[(b)]
The boundary values of $\Psi(\zeta)$ on $\Upsilon \setminus \{0\}$
are related by the conditions
    \begin{align}
        \label{RHP Psi2: b1}
        \Psi_+(\zeta) &= \Psi_-(\zeta)
            \begin{pmatrix}
                1 & 0 \\
                -i & 1
            \end{pmatrix},
            \qquad \mbox{for $\zeta\in\Upsilon_1\cup\Upsilon_{2n+1}$,}
        \\[1ex]
        \label{RHP Psi2: b2}
            \Psi_+(\zeta) &= \Psi_-(\zeta)
            \begin{pmatrix}
                1 & -i \\
                0 & 1
            \end{pmatrix},
            \qquad \mbox{for $\zeta\in\Upsilon_{2n+2}\cup\Upsilon_{4n+2}$,}
    \end{align}
\item[(c)] $\Psi$ has the following behavior at infinity:
    \begin{equation}\label{RHP Psi2: c}
        \Psi(\zeta)=(I+\Psi_\infty\frac{1}{\zeta}+\bigO(\zeta^{-2}))e^{-i\Theta(\zeta)\sigma_3},
        \qquad \mbox{as $\zeta\to\infty$,}\end{equation}
        where
        \begin{equation}\label{def hattheta}
        \Theta(\zeta;x,\tau_1,\ldots,\tau_{n-1})=\frac{1}{4n+2}(2\zeta)^{2n+1}+
\sum_{j=1}^{n-1}\frac{(-1)^{j+1}
        \tau_{j}}{4j+2}(2\zeta)^{2j+1}+x\zeta.
    \end{equation}
\item[(d)] Near the origin,
    \begin{equation}\label{RHP Psi2: d}
        \Psi(\zeta)=
        \begin{cases}
            \bigO\begin{pmatrix}
                |\zeta|^{\frac{1}{2}} & |\zeta|^{-\frac{1}{2}} \\
                |\zeta|^{\frac{1}{2}} & |\zeta|^{-\frac{1}{2}}
            \end{pmatrix}
                     \begin{pmatrix} 1 & 0 \\ \pm i & 1 \end{pmatrix},
                & \mbox{as $\zeta\to 0$, $\zeta \in S_1$ for ``+'',
                                    $\zeta  \in S_3$ for ``-'',}
            \\[3ex]
            \bigO\begin{pmatrix}
                |\zeta|^{\frac{1}{2}} & |\zeta|^{-\frac{1}{2}} \\
                |\zeta|^{\frac{1}{2}} & |\zeta|^{-\frac{1}{2}}
            \end{pmatrix},
                & \mbox{as $\zeta\to 0,\, \zeta \in S_2$,}
            \\[3ex]
            \bigO\begin{pmatrix}
                |\zeta|^{-\frac{1}{2}} & |\zeta|^{\frac{1}{2}} \\
                |\zeta|^{-\frac{1}{2}} & |\zeta|^{\frac{1}{2}}
            \end{pmatrix},
                & \mbox{as $\zeta\to 0,\, \zeta \in S_4$.}
        \end{cases}
    \end{equation}
\end{itemize}

\begin{remark}
The conditions (a,b) are a special case of the conditions
(a,b) of the RH problem of the previous section, corresponding to our
choice of the Stokes multipliers. The condition (c) remains the same.
For our choice of Stokes multipliers, it was proven \cite{FN, FIKN} that $\Psi$ can be written near $0$ in the form:
\begin{equation*}
\Psi(\zeta)=E(\zeta)\begin{pmatrix}\zeta^{-\frac{1}{2}}&0\\
\frac{1}{\pi}\zeta^{\frac{1}{2}}\ln\zeta &\zeta^{\frac{1}{2}}
\end{pmatrix}A_j,\qquad\mbox{ for $\zeta\in S_j$,}
\end{equation*}
with
\begin{align*}&A_1=\begin{pmatrix}-i&-1\\1+ip&p\end{pmatrix}, & A_2=\begin{pmatrix}0&-1\\1&p\end{pmatrix}, \\ & A_3=\begin{pmatrix}i&-1\\1-ip&p\end{pmatrix}, & A_4=\begin{pmatrix}i&0\\1-ip&-i\end{pmatrix},\end{align*}
for some $p\in\mathbb C$, and with $E$ analytic at $0$.
The condition (\ref{RHP Psi2: d}) follows from this formula.
\end{remark}

\begin{remark}
Using Liouville's theorem, one verifies directly that this RH problem, if solvable, has a unique solution.
\end{remark}

From now on, we write $\Psi$ for the solution to this particular case of the RH problem for $\P2n$, instead of the more general RH problem with arbitrary Stokes multipliers.
Starting from this RH problem, we will now prove Theorem \ref{theorem: q}.
First, we will prove reality and smoothness of $q$
following \cite{CKV}, where this fact was proven in the case $n=1$.
Afterwards we will analyze the RH problem asymptotically using the Deift and
Zhou steepest descent method in order to obtain asymptotics for $q$ as $x\to\pm\infty$.

\subsection{Smoothness and reality of $q$}\label{section Painleve-smoothness}

Let us first prove that the Painlev\'e solution $q$ corresponding to our RH problem is real.
If $\Psi$ solves the RH problem given in Section \ref{section: RHP}, then it is easily verified that also $\sigma_1\overline{\Psi(\overline\zeta)}\sigma_1$ and $\sigma_1\Psi(-\zeta)\sigma_1$ are solutions of the RH problem, with $\sigma_1=\begin{pmatrix}0&1\\1&0\end{pmatrix}$. Because of the uniqueness we have
\begin{equation}\label{symmetry real}
\Psi(\zeta)=\sigma_1\overline{\Psi(\overline\zeta)}\sigma_1=\sigma_1\Psi(-\zeta)\sigma_1.
\end{equation}
Consequently, the matrix $\Psi_\infty$ appearing in (\ref{RHP Psi2: c}) satisfies
\[\Psi_\infty=\sigma_1\overline{\Psi_\infty}\sigma_1=-\sigma_1\Psi_\infty\sigma_1.\]
By (\ref{qPsi}), this implies the reality of $q$.

\medskip

We will now show that $q$ has no poles for real values of $x,\tau_1, \ldots, \tau_{n-1}$.
In view of (\ref{qPsi}), this is equivalent to the solvability of
the RH problem of Section \ref{section: RHP}
for all real values of the parameters.
A proof of solvability is based on the so-called vanishing lemma.
This technique was developed in \cite{FokasMuganZhou, FokasZhou} and is explained in detail in \cite{FIKN}.

\begin{lemma}{\bf (vanishing lemma)}\label{VL}
 Let $\Psi_0$ satisfy the conditions (a), (b), and (d) of the RH problem for $\Psi$ in Section \ref{section: RHP}, and let the asymptotic condition at infinity be replaced by the homogeneous condition
\begin{equation}\label{vanishing condition}
\Psi_0(\zeta)e^{i\Theta(\zeta)\sigma_3}=\bigO(\zeta^{-1}), \qquad \mbox{ as $\zeta\to\infty$.}
\end{equation}
Then $\Psi_0\equiv 0$ is the only solution to this RH problem.
\end{lemma}
The proof of this vanishing lemma was given in \cite[Proposition 2.5]{CKV} in the case where $n=1$ and $\Theta(\zeta;x)=\frac{4}{3}\zeta^3+x\zeta$ (for general $\alpha>-\frac{1}{2}$). For arbitrary $n$ and $\Theta$ given by (\ref{def Theta}) however, the proof remains exactly the same. It should be noted that the proof does not remain valid for any other choice of Stokes multipliers.

\medskip

By a standard argument \cite[Section 5.3]{DKMVZ2}, \cite{FIKN, KMM},
it follows from Lemma \ref{VL} that the RH problem of
Section \ref{section: RHP} is solvable.

\subsection{Asymptotics for $q$ at $+\infty$ for $n$ odd}\label{section as+}

We will now perform an asymptotic analysis of the RH problem for the special smooth solution of $\P2n$ as $x\to +\infty$. This steepest descent analysis shows many similarities with the analysis we did in Section \ref{section large gap} in order to obtain large gap asymptotics for the Fredholm determinant as $s\to -\infty$.

\subsubsection{Rescaling and normalization of the RH problem}

\begin{figure}[t]
\begin{center}
    \setlength{\unitlength}{1truemm}
    \begin{picture}(100,48.5)(0,2.5)
        \put(72,27.5){\small $\wt S_1$}
        \put(49,49.5){\small $\wt S_2$}
        \put(27.5,27.5){\small $\wt S_3$}
        \put(49,5.5){\small $\wt S_4$}

        \put(77,42.5){\small $\widetilde\Upsilon_1$}
        \put(17.5,42.5){\small $\widetilde\Upsilon_{2n+1}$}
        \put(17.5,10){\small $\widetilde\Upsilon_{2n+2}$}
        \put(77,10){\small $\widetilde\Upsilon_{4n+2}$}

        \put(50,32.5){\thicklines\vector(0,1){.0001}}
         \put(50,22.5){\thicklines\vector(0,-1){.0001}}
        \put(47,28){\small 0}
        \put(50,27.5){\thicklines\circle*{.8}}

        \put(50,20.5){\line(0,1){14}}
        \put(15,3){\line(2,1){35}}
        \put(15,52){\line(2,-1){35}}
        \put(50,34.5){\line(2,1){35}}
        \put(50,20.5){\line(2,-1){35}}
        \put(70,44.5){\thicklines\vector(2,1){.0001}}
        \put(70,10.5){\thicklines\vector(2,-1){.0001}}
        \put(30,44.5){\thicklines\vector(-2,1){.0001}}
        \put(30,10.5){\thicklines\vector(-2,-1){.0001}}
    \end{picture}
    \caption{Deformed contour $\widetilde\Upsilon$}
    \label{figure: deformed contour}
\end{center}
\end{figure}

\begin{figure}[t]
\begin{center}
\includegraphics[scale=0.4]{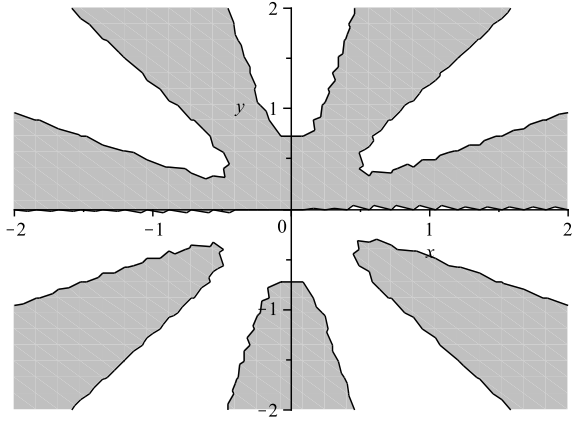}
\end{center}
\caption{Contour plot of $\Im\Theta(x^{\frac{1}{2n}}\zeta)$ for $x>0$ large and $n=3$. The shaded
areas indicate where $\Im\Theta(x^{\frac{1}{2n}}\zeta)>0$. The contour $\widetilde\Upsilon$ should lie in the shaded region in the upper half plane and in the white region in the lower half plane. \label{figure: contourplot}}
\end{figure}

Let us define
\begin{equation}\label{def T}
T(\zeta)=\Psi(x^{\frac{1}{2n}}\zeta)
\exp\{i\Theta(x^{\frac{1}{2n}}\zeta)\sigma_3\}.
\end{equation}
First of all, note that the multiplication on the right with this exponential is the most obvious way to normalize the RH problem at infinity, see (\ref{RHP Psi2: c}).
Secondly, the rescaling with the factor $x^{\frac{1}{2n}}$ is needed in order to balance the leading order term and the linear term of $\Theta$ as $x\to +\infty$. Indeed we have that
\begin{multline}\label{rescaledTheta}
\Theta(x^{\frac{1}{2n}}\zeta;x,\tau_1, \ldots , \tau_{n-1})=\\ x^{\frac{2n+1}{2n}}\left(\frac{1}{4n+2}(2\zeta)^{2n+1}+\zeta+\bigO(x^{-\frac{1}{2n}})\right), \qquad \mbox{ as $x\to +\infty$.}
\end{multline}
Balancing those terms is crucial for our analysis. The coefficient of $\zeta$ in the above expression will turn out to have a direct influence on the leading order asymptotics for $q(x)$ as $x\to +\infty$.
Let us deform the jump contour $\Upsilon$ (see Remark \ref{remark freedom contour}) to a contour $\widetilde\Upsilon$ as shown in Figure \ref{figure: deformed contour}, where we let the curves  $\widetilde\Upsilon_1, \widetilde\Upsilon_{2n+1}$ and $\widetilde\Upsilon_{2n+2},\widetilde\Upsilon_{4n+2}$ coincide with the imaginary axis on an interval $(-i\de,i\de)$ for a small fixed
$\de>0$, and we let $\tilde\Upsilon_j$ be parallel to $\Upsilon_j$ away from $(-i\de,i\de)$.
 Away from the origin, we should choose the contour so that it lies in the interior of the region where
\begin{align}\label{Im}&\Im\left(\frac{1}{4n+2}(2\zeta)^{2n+1}+\zeta\right)>0, \qquad\mbox{ as $\zeta\in\widetilde\Upsilon$, $\Im\zeta>0$},\\
\label{Im2}&\Im\left(\frac{1}{4n+2}(2\zeta)^{2n+1}+\zeta\right)<0, \qquad\mbox{ as $\zeta\in\widetilde\Upsilon$, $\Im\zeta<0$}.\end{align}
One verifies directly that this is the case for small $\delta$, which is also visible from Figure \ref{figure: contourplot}. The deformation near the origin is not crucial but simplifies the construction of a local parametrix later on. The condition (\ref{Im})-(\ref{Im2}) is essential in order to have exponentially decaying jump matrices.
We obtain the new $\wt T$ corresponding to $\wt\Upsilon$ in a straightforward way by continuing $T$ analytically through $\Upsilon$.
Namely, if $J_1(\ze)$, $J_{2n+1}(\ze)$, $J_{2n+2}(\ze)$, $J_{4n+2}(\ze)$,
are the jumps of $T$ on $\Upsilon$, i.e. $T_+=T_-J$, and $\wt\Om_j$
for each $j=1,2n+1,2n+2,4n+2$ is the region bounded by $\Upsilon_j$,
$\wt\Upsilon_j$, and $(0,i\de(-1)^{j+1})$, we set
\be\label{def wtT}
\wt T(\ze)=\begin{cases}
T, &\ze\in S_1, S_3, \wt S_2, \wt S_4\cr
TJ_1^{-1}, &\ze\in\wt\Om_1\cr
TJ_{2n+1},&\ze\in\wt\Om_{2n+1}\cr
TJ_{2n+2}^{-1},&\ze\in\wt\Om_{2n+2}\cr
TJ_{4n+2},&\ze\in\wt\Om_{4n+2}
\end{cases}
\ee
We then have
\subsubsection*{RH problem for $\wt T$}
\begin{itemize}
\item[(a)] $\wt T$ is analytic in $\mathbb C\setminus\widetilde\Upsilon$.
\item[(b)] The boundary values of $\wt T$ on $\widetilde\Upsilon \setminus \{0\}$
       satisfy the jump relations
    \begin{align}
        \label{RHP T: b1+}
        \wt T_+(\zeta) &= \wt T_-(\zeta)
            \begin{pmatrix}
                1 & 0 \\
                -ie^{2i\Theta(x^{\frac{1}{2n}}\zeta)} & 1
            \end{pmatrix},
            \qquad \mbox{for $\zeta\in\left(\widetilde\Upsilon_1\cup\widetilde\Upsilon_{2n+1}\right)\setminus[0,i\delta]$,}
        \\[1ex]
        \label{RHP T: b2+}
            \wt T_+(\zeta) &= \wt T_-(\zeta)
            \begin{pmatrix}
                1 & -ie^{-2i\Theta(x^{\frac{1}{2n}}\zeta)} \\
                0 & 1
            \end{pmatrix},
            \qquad \mbox{for $\zeta\in\left(\widetilde\Upsilon_{2n+2}\cup\widetilde\Upsilon_{4n+2}\right)\setminus[-i\delta,0]$,}\\[1ex]
        \label{RHP T: b3}
            \wt T_+(\zeta) &= \wt T_-(\zeta)
            \begin{pmatrix}
                1 & -2ie^{-2i\Theta(x^{\frac{1}{2n}}\zeta)} \\
                0 & 1
            \end{pmatrix},
            \qquad \mbox{for $\zeta\in(-i\delta,0)$,}\\[1ex]
        \label{RHP T: b4+}
            \wt T_+(\zeta) &= \wt T_-(\zeta)
            \begin{pmatrix}
                1 & 0 \\
                -2ie^{2i\Theta(x^{\frac{1}{2n}}\zeta)} & 1
            \end{pmatrix},
            \qquad \mbox{for $\zeta\in(0,i\delta)$.}
    \end{align}
\item[(c)]As $\zeta\to\infty$, we have
    \begin{equation}\label{RHP T: c+}
        \wt T(\zeta)=I+\frac{\Psi_\infty}{x^{\frac{1}{2n}}\zeta}+\bigO(\zeta^{-2}).
        \end{equation}
\item[(d)] Near the origin,
    \begin{equation}\label{RHP T: d}
         \wt T(\zeta)\begin{pmatrix}
                1&0\\ \mp i&1
            \end{pmatrix}=
            \bigO\begin{pmatrix}
                |\zeta|^{\frac{1}{2}} & |\zeta|^{-\frac{1}{2}} \\
                |\zeta|^{\frac{1}{2}} & |\zeta|^{-\frac{1}{2}}
            \end{pmatrix},
                \qquad\mbox{as $\zeta\to 0,\, \pm\Re\zeta >0 $.}
    \end{equation}
\end{itemize}


\medskip

It follows directly from (\ref{rescaledTheta}) and (\ref{Im}), (\ref{Im2}) that, as $x\to +\infty$, the jump matrices for $\wt T$ tend to the identity matrix exponentially fast on the jump contour $\widetilde\Upsilon$ except near the origin, say in a disk $U_0$ of radius $\delta$ centered at $0$.
Moreover, the behaviour of the jump matrices for a fixed $x$ as $\ze\to\infty$
ensures that the transformation $T\rightarrow\wt T$ preserves
the condition (c).

\subsubsection{Local parametrix near the origin}\label{section + local}
We construct a local parametrix in $U_0$ using Bessel functions. The model RH problem we use is different from the one we used in Section \ref{section large gap}, although these problems are related.

\subsubsection*{Bessel model RH problem}
Let us
define the functions $M_1$, $M_2$ as follows:
\begin{align}
&M_1(\lambda)=\frac{1}{2}e^{-i\pi\sigma_3/4}\begin{pmatrix}1&i\\i&1\end{pmatrix}
\sqrt{\frac{\pi\lambda}{2}}
\begin{pmatrix}-iH_{1}^{(2)}(\lambda)&H_{1}^{(1)}(\lambda)\\
-iH_{0}^{(2)}(\lambda)& H_{0}^{(1)}(\lambda)\end{pmatrix},
&\mbox{for $\lambda\in\mathbb C\setminus \mathbb R^-$,}\\
&M_2(\lambda)=\frac{1}{2}e^{-i\pi\sigma_3/4}\begin{pmatrix}1&i\\i&1\end{pmatrix}
\sqrt{\frac{\pi\lambda}{2}}
\begin{pmatrix}iH_{1}^{(1)}(\lambda e^{-i\pi})&H_{1}^{(2)}(\lambda e^{-i\pi})\\
-iH_{0}^{(1)}(\lambda e^{-i\pi})& -H_{0}^{(2)}(\lambda e^{-i\pi})\end{pmatrix}
,&\mbox{for $\lambda\in\mathbb C\setminus \mathbb R^+$,}
\end{align}
where $H_0^{(1)}$ and $H_0^{(2)}$ again denote Hankel functions,
with their branch cuts on the negative real axis.
We then have the relations:
\begin{align}
            \label{RHP M: b1}
            M_2(\lambda) &= M_1(\lambda)
                \begin{pmatrix}
                    1 & 0\\
                    -2i & 1
                \end{pmatrix},&&\mbox{ for $\lambda$ in the upper half-plane,}\\[1ex]
            \label{RHP M: b2}
            M_1(\lambda) &= M_2(\lambda)
                \begin{pmatrix}
                    1 & -2i\\
                    0 & 1
                \end{pmatrix},&&\mbox{ for $\lambda$ in the lower half-plane.}
        \end{align}
Using the asymptotic properties for the Hankel functions, one verifies that
$M_k$ has the following asymptotics for $\lambda\to\infty$:
\begin{multline}\label{RHP M: c}
            M_k(\lambda)=\left[I+\frac{1}{8\lambda}
            \begin{pmatrix}-i&-2i\\2i&i\end{pmatrix}+\bigO(\lambda^{-2})\right]e^{-i\lambda\sigma_3},\\ \qquad
                \mbox{as $\lambda\to\infty$ in sector $\widehat S_k$,}
        \end{multline}
with \begin{align}\label{sectorsM}
&\widehat S_1=\{\lambda: -\pi+\epsilon <\arg\lambda< \pi -\epsilon\},\\
&\widehat S_2=\{\lambda: \epsilon <\arg\lambda< 2\pi-\epsilon\}.
\end{align}
We will use $M_1$ and $M_2$ as 'model functions' to construct a local approximation to the RH solution $\wt T$ near the origin.
\subsubsection*{Construction of the parametrix}
We define the parametrix $P$ as follows:
\begin{equation}\label{def P}
P(\zeta)=\begin{cases}\begin{array}{ll}M_1(x^{\frac{2n+1}{2n}}f(\zeta))e^{i\Theta(x^{\frac{1}{2n}}\zeta)\sigma_3},&\mbox{ for $\Re\zeta>0$, $\zeta\in U_0$,}\\
M_2(x^{\frac{2n+1}{2n}}f(\zeta))e^{i\Theta(x^{\frac{1}{2n}}\zeta)\sigma_3},&\mbox{ for $\Re\zeta<0$, $\zeta\in U_0$.}\end{array}\end{cases}
\end{equation}
where
\begin{equation}
f(\zeta)=x^{-\frac{2n+1}{2n}}\Theta(x^{\frac{1}{2n}}\zeta).
\end{equation}
From (\ref{rescaledTheta}), we have that
\begin{equation}\label{f +}
f(0)=0, \qquad f'(0)=1.
\end{equation}
Using the RH conditions for $M_1$ and $M_2$
we obtain (for sufficiently small $U_0$):
\subsubsection*{RH problem for $P$}
\begin{itemize}
    \item[(a)] $P(\zeta)$ is analytic in $U_0\setminus\widetilde\Upsilon$.
    \item[(b)] $P$ satisfies the following jump relations on
    $U_0\cap \widetilde\Upsilon$:
    \begin{align}\label{RHP P2:b1}
        P_+(\zeta)&=P_-(\zeta)
            \begin{pmatrix}
                1 & 0 \\
                -2ie^{2i\Theta(x^{\frac{1}{2n}}\zeta)} & 1
            \end{pmatrix},&& \mbox{for $\zeta\in U_0\cap [0,i\delta]$,}\\[1ex]
            \label{RHP P2:b2}
        P_+(\zeta)&=P_-(\zeta)
            \begin{pmatrix}
                1 & -2i e^{-2i\Theta(x^{\frac{1}{2n}}\zeta)}\\
                0 & 1
            \end{pmatrix},&& \mbox{for
            $\zeta\in U_0\cap [-i\delta,0]$.}
    \end{align}
    \item[(c)] At the boundary $\partial U_0$ of $U_0$ as $x\to +\infty$,
    \begin{equation}\label{RHP P2:c}
        P(\zeta)=I+\frac{1}{8x^{\frac{2n+1}{2n}}f(\zeta)}
            \begin{pmatrix}-i&-2i\\2i&i\end{pmatrix}+\bigO\left(x^{-\frac{2n+1}{n}}\right),
    \end{equation}
    \item[(d)] Near the origin,
    \begin{equation}\label{RHP hatP+:d}
    P(\zeta)\begin{pmatrix}
                1&0\\\mp i&1
            \end{pmatrix}=
            \bigO\begin{pmatrix}
                |\zeta|^{\frac{1}{2}} & |\zeta|^{-\frac{1}{2}} \\
                |\zeta|^{\frac{1}{2}} & |\zeta|^{-\frac{1}{2}}
            \end{pmatrix},
                \qquad\mbox{as $\zeta\to 0,\, \pm\Re\zeta >0 $.}
                \end{equation}
\end{itemize}

The local condition (d) follows from the behavior of Hankel functions
as $\lambda\to 0$.

\medskip

Note that the outside parametrix is simply given by the identity matrix
since, away from $0$, the jump matrices for $\wt T$ tend to $I$ everywhere.

\subsubsection{Final RH problem and asymptotics for $q(x)$ as $x\to +\infty$}

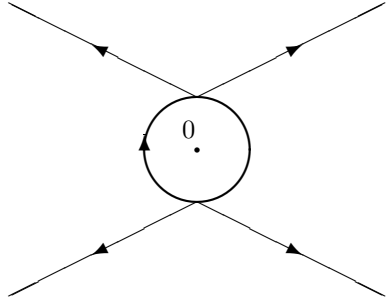
\begin{figure}[t]
\begin{center}
    \setlength{\unitlength}{1truemm}
    \begin{picture}(100,48.5)(0,2.5)
        \put(50,27.5){\thicklines\circle{14}}
        \put(43,29.5){\thicklines\vector(0,1){.0001}}

        \put(48,29){\small 0}
        \put(50,27.5){\thicklines\circle*{.8}}

        \put(50,34.5){\line(2,1){25}}
        \put(50,34.5){\line(-2,1){25}}
        \put(50,20.5){\line(-2,-1){25}}
        \put(50,20.5){\line(2,-1){25}}
        \put(64,41.5){\thicklines\vector(2,1){.0001}}
        \put(64,13.5){\thicklines\vector(2,-1){.0001}}
        \put(36,41.5){\thicklines\vector(-2,1){.0001}}
        \put(36,13.5){\thicklines\vector(-2,-1){.0001}}
    \end{picture}
    \caption{Contour $\Sigma_R$}
    \label{figure: SigmaR-2+}
\end{center}
\end{figure}

Let us define
\begin{align}\label{def R1-2}
&R(\zeta)=\wt T(\zeta)P(\zeta)^{-1}&&\mbox{ for $\zeta\in U_0$,}\\
&\label{def R2-2}R(\zeta)=\wt T(\zeta),&&\mbox{ for $\zeta\in \mathbb C\setminus U_0$.}
\end{align}
From the behavior of $\wt T$ and $P$ near the origin it follows that $R$ has
a removable singularity at $0$. The function
$R$ satisfies a RH problem with jump matrices which are uniformly close
to the identity matrix.
\subsubsection*{RH problem for $R$}
\begin{itemize}
\item[(a)] $R$ is analytic in $\mathbb C\setminus \Sigma_R$, with $\Sigma_R$ as shown in Figure
\ref{figure: SigmaR-2+}.
\item[(b)] $R$ satisfies the jump relations $R_+=R_-v_R$ on $\Sigma_R$, with
\begin{align}
&v_R(\zeta)=P(\zeta),&&\mbox{ for $\zeta\in \partial U_0$,}\\
&v_R(\zeta)=v_T(\zeta),&&\mbox{ for $\zeta\in\Sigma_R\setminus\overline{U_0}$.}
\end{align}
\item[(c)] $R(\zeta)\to I$ as $\zeta\to\infty$.
\end{itemize}
Outside $\overline U_0$, we have already observed that
\[v_R(\zeta)=v_T(\zeta)=I+\bigO(e^{-cx^{\frac{2n+1}{2n}}}),\qquad\mbox{ as $x\to +\infty$.}\]
For $\zeta\in\partial U_0$, we obtain from (\ref{RHP P2:c}) that in the limit as $x\to +\infty$,
\begin{equation}\label{expansion JR2}
v_R(\zeta)=I+\frac{v_1(\zeta)}{x^{\frac{2n+1}{2n}}}
+\bigO\left(x^{-\frac{2n+1}{n}}\right),
\qquad v_1(\zeta)=\frac{1}{8f(\zeta)}
            \begin{pmatrix}-i&-2i\\2i&i\end{pmatrix}.
\end{equation}
This situation is similar to the one in Section \ref{section: large gap final}, and we conclude in exactly the same way that
\begin{equation}\label{R3}
R(\zeta)=I+\frac{R_1(\zeta)}{x^{\frac{2n+1}{2n}}}+\bigO(x^{-\frac{2n+1}{n}}), \qquad \mbox{ as $x\to +\infty$},
\end{equation}
with
\[R_1(\zeta)=\frac{1}{\zeta}\Res(v_1;0)=
\frac{1}{8\zeta}
            \begin{pmatrix}-i&-2i\\2i&i\end{pmatrix},
\qquad\mbox{ for $\zeta\in\mathbb C\setminus \overline U_0$.}\]
Using the fact that $\wt T(\zeta)=R(\zeta)$ for $\zeta\in\mathbb C\setminus \overline U_0$ and (\ref{RHP T: c+}), we obtain
\begin{equation}
\Psi_\infty=\frac{1}{8x}
            \begin{pmatrix}-i&-2i\\2i&i\end{pmatrix}+\bigO(x^{-\frac{4n+1}{2n}})\qquad \mbox{ as $x\to +\infty$},
\end{equation}
so that by (\ref{qPsi}),
\begin{equation}\label{as q+}
q(x;\tau_1, \ldots , \tau_{n-1})=\frac{1}{2x}+\bigO(x^{-\frac{4n+1}{2n}})\qquad \mbox{ as $x\to +\infty$},
\end{equation}
which proves part of Theorem \ref{theorem: q} (equation
(\ref{asymptotics q+}).

For later use in Section \ref{section Painleve}, consider
\be
\label{defdefE}
E(\ze)\equiv
{1\over\sqrt{2}}\begin{pmatrix} 1 & 0 \cr -q(x) & 1 \end{pmatrix}
\zeta^{-\frac{1}{4}\sigma_3}\begin{pmatrix}1&1\\-1&1\end{pmatrix}
\Psi(i\zeta^{\frac{1}{2}})
e^{-\frac{1}{4}\pi i\sigma_3}
 \begin{pmatrix}1&-{1\over 2\pi i}\ln\ze\\0 &1\end{pmatrix}
\ee
with the principal branch of the root and the logarithm (the branch cut
is $(-\infty,0)$).
We will need to know the asymptotics of
$E_{0,11}=\lim_{\ze\searrow 0}E_{11}(\ze,x,\tau_1,\dots,\tau_{n-1})$
as $x\to+\infty$. The $\ze$-limit here is
taken so that $u=i\ze^{1/2}$ tends to zero
along $(0,+i\infty)$.
We have
\begin{equation}\label{defE}
E_{0,11}=\frac{e^{-\frac{i\pi}{4}}}{\sqrt 2}\lim_{\zeta\searrow 0}\zeta^{-1/4}\left[\begin{pmatrix}1&1\\-1&1\end{pmatrix}\Psi(i\zeta^{1/2})\right]_{11}.
\end{equation}
From (\ref{def wtT}), (\ref{def R1-2}), and (\ref{R3}), we obtain
as $x\to +\infty$
\begin{equation*}
E_{0,11}=\frac{e^{-\frac{i\pi}{4}}}{\sqrt 2}\lim_{\zeta\searrow 0}
\zeta^{-1/4}\left[\begin{pmatrix}1&1\\-1&1\end{pmatrix}
\left(I+\bigO(x^{-\frac{2n+1}{2n}})\right)
P_-(ix^{-\frac{1}{2n}}\zeta^{1/2})
\begin{pmatrix}1&0\\-i&1\end{pmatrix}\right]_{11}.
\end{equation*}
Using the definition of the parametrix (\ref{def P}) and the
expansions for Hankel functions near the origin,
we finally find
\begin{equation}\label{as E011+}
E_{0,11}=e^{\frac{-i\pi}{4}}\sqrt{\pi x}(1+o(1)), \qquad \mbox{ as $x\to +\infty$.}
\end{equation}

\subsubsection{Asymptotics at $+\infty$ for other solutions of the Painlev\'e II hierarchy}\label{section general+}

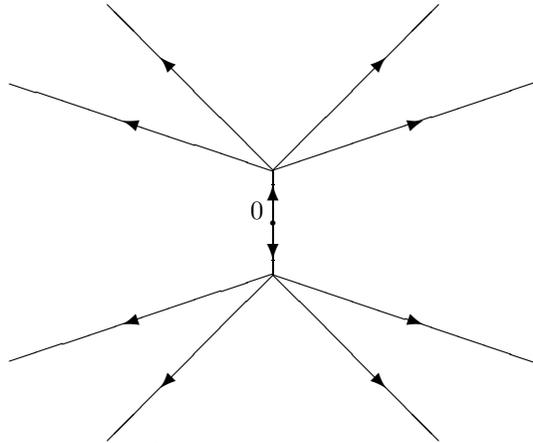
\begin{figure}[t]
\begin{center}
    \setlength{\unitlength}{1truemm}
    \begin{picture}(100,48.5)(0,2.5)

        \put(50,32.5){\thicklines\vector(0,1){.0001}}
         \put(50,22.5){\thicklines\vector(0,-1){.0001}}
        \put(47,28){\small 0}
        \put(50,27.5){\thicklines\circle*{.8}}

        \put(50,34.5){\line(-1,1){22}}
        \put(50,20.5){\line(-1,-1){22}}
        \put(50,20.5){\line(0,1){14}}
        \put(15,9){\line(3,1){35}}
        \put(15,46){\line(3,-1){35}}
        \put(50,34.5){\line(3,1){35}}
        \put(50,20.5){\line(3,-1){35}}
        \put(50,34.5){\line(1,1){22}}
        \put(50,20.5){\line(1,-1){22}}

        \put(70,41){\thicklines\vector(3,1){.0001}}
        \put(70,14){\thicklines\vector(3,-1){.0001}}
        \put(30,41){\thicklines\vector(-3,1){.0001}}
        \put(30,14){\thicklines\vector(-3,-1){.0001}}

        \put(65,49.5){\thicklines\vector(1,1){.0001}}
        \put(65,5.5){\thicklines\vector(1,-1){.0001}}
        \put(35,49.5){\thicklines\vector(-1,1){.0001}}
        \put(35,5.5){\thicklines\vector(-1,-1){.0001}}

    \end{picture}
    \caption{Deformed contour $\widetilde\Upsilon$ for solutions with $4$ non-zero Stokes multipliers}
    \label{figure: contour general Stokes}
\end{center}
\end{figure}

The special solution to $\P2n$ which we considered, i.e.\ the one with Stokes multipliers
\begin{align}\label{Stokesmultipliers-3}
&s_1=s_{2n+1}=-i,\\
&s_2=s_3= \cdots =s_{2n}=0,
\end{align}
is not the only solution which has the asymptotic behavior given by (\ref{as q+}).
For $n$ odd, any solution for which the Stokes multipliers satisfy the conditions
\begin{align}\label{Stokesmultipliers-4}
&s_2=s_{4}=\cdots =s_{2n}=0,\\
\label{Stokesmultipliers-5}&s_1+s_3+ \cdots +s_{2n+1}=-2i,
\end{align}
shares this behavior at $+\infty$. These facts can be proved in a very similar way to the proof we gave for our special solution $q$. The main difference in the steepest descent analysis is a deformation of the jump contour for $T$ to a contour similar to the one shown in Figure \ref{figure: contour general Stokes}.
If one chooses the curves so that they coincide near $0$ as indicated in Figure \ref{figure: contour general Stokes}, the local parametrix near $0$ can be constructed in exactly the same way as in Section \ref{section + local}. Furthermore, if the contour is chosen so that conditions (\ref{Im}), (\ref{Im2})
hold, the jump matrices for $R$ still show uniform decay.

%

\medskip

If some of the 'even' Stokes multipliers $s_2, s_4, \ldots $ are non-zero, it might still be possible to construct the local parametrix, but there is no way to deform the jump contour in such a way that (\ref{Im}) remains valid. In the spirit of the steepest descent method for Painlev\'e equations, this means (see \cite{FIKN} for several examples of this procedure) that one needs to construct a so-called $g$-function, which leads typically to a RH problem for which the outside parametrix is not the identity matrix, as it was in our analysis. The leading order of the Painlev\'e transcendent will then be determined by this outside parametrix. This provides a non-rigorous but strong heuristic argument to believe that only the solutions with Stokes multipliers satisfying (\ref{Stokesmultipliers-4})--(\ref{Stokesmultipliers-5}) share the asymptotic behavior at $+\infty$ given by (\ref{as q+}).

\subsection{Asymptotics for $q$ at $-\infty$ for $n$ odd}
We will now perform a similar analysis for $q(x)$ as $x\to -\infty$. The main difference with the analysis at $+\infty$ is that the normalization of the RH problem needs to be done in a less obvious way, using a $g$-function.

\subsubsection{Rescaling of the RH problem and deformation of the jump contour}

\begin{figure}[t]
\begin{center}
    \setlength{\unitlength}{1truemm}
    \begin{picture}(100,48.5)(0,2.5)
        \put(72,27.5){\small $\widehat S_1$}
        \put(49,49.5){\small $\widehat S_2$}
        \put(27.5,27.5){\small $\widehat S_3$}
        \put(49,5.5){\small $\widehat S_4$}

        \put(77,42.5){\small $\widehat \Upsilon_1$}
        \put(17.5,42.5){\small $\widehat \Upsilon_{2n+1}$}
        \put(17.5,10){\small $\widehat \Upsilon_{2n+2}$}
        \put(77,10){\small $\widehat \Upsilon_{4n+2}$}

        \put(45,27.5){\thicklines\vector(-1,0){.0001}}
         \put(55,27.5){\thicklines\vector(1,0){.0001}}
        \put(48,29){\small 0}
        \put(50,27.5){\thicklines\circle*{.8}}

        \put(43,27.5){\line(1,0){14}}
        \put(43,27.5){\line(-2,1){35}}
        \put(43,27.5){\line(-2,-1){35}}
        \put(57,27.5){\line(2,1){35}}
        \put(57,27.5){\line(2,-1){35}}
        \put(77,37.5){\thicklines\vector(2,1){.0001}}
        \put(77,17.5){\thicklines\vector(2,-1){.0001}}
        \put(23,37.5){\thicklines\vector(-2,1){.0001}}
        \put(23,17.5){\thicklines\vector(-2,-1){.0001}}
    \end{picture}
    \caption{Deformed contour $\widehat\Upsilon$}
    \label{figure: deformed contour-}
\end{center}
\end{figure}

Let us first make a rescaling:
\begin{equation}\label{def T-}
T(\zeta)=\Psi(|x|^{\frac{1}{2n}}\zeta),
\end{equation}
and choose a deformed jump contour $\widehat\Upsilon$ as indicated in Figure \ref{figure: deformed contour-}. We let $\widehat\Upsilon_{4n+2}$ and $\widehat\Upsilon_1$ coincide along $[0,\zeta_0]$ and $\widehat\Upsilon_{2n+1}$ and $\widehat\Upsilon_{2n+2}$ coincide along $[-\zeta_0,0]$. Here $\zeta_0$ is real and positive, and we will determine the precise value of $\zeta_0$ later on.
Away from $[-\zeta_0, \zeta_0]$, we take each ray in the contour to make an angle of $\frac{\pi}{4n+2}$ with the real line. It will become clear later why this deformation is convenient.
We now define a matrix $\widehat T$ in terms of $T$, in a similar way
as $\wt T$ was defined in (\ref{def wtT}). Namely, in the region bounded by
$\Upsilon_1$, $\widehat\Upsilon_1$, and $(0,\ze_0)$, as well as in the region
bounded by
$\Upsilon_{2n+1}$, $\widehat\Upsilon_{2n+1}$, and $(-\ze_0,0)$,
the function $\hat T$ is defined by extending $T$ analytically from $S_2$.
In the region bounded by
$\Upsilon_{4n+2}$, $\widehat\Upsilon_{4n+2}$, and $(0,\ze_0)$,
as well as in the region bounded by
$\Upsilon_{2n+2}$, $\widehat\Upsilon_{2n+2}$, and $(-\ze_0,0)$,
the function $\widehat T$ is defined by extending $T$ analytically from $S_4$.
In the regions $\widehat S_1$,  $\widehat S_3$, $S_2$ and $S_4$, we set
$\widehat T=T$.

\subsubsection{Construction of the $g$-function and normalization of the RH problem}

In the previous section where we computed asymptotics at $+\infty$, we used the obvious way to normalize the RH problem, namely, multiplying on the right with $e^{i\Theta(|x|^{\frac{1}{2n}}\zeta)\sigma_3}$. Here we could do this as well, but it would not lead to a RH problem with decaying jump matrices. This is a consequence of the fact that the topology of the set $\{\zeta: \Im \Theta(|x|^{\frac{1}{2n}}\zeta)>0\}$ for negative $x$ is different
from that in Figure \ref{figure: contourplot}. We will deal with this problem by replacing $\Theta$ by a function $g$ behaving like $\Theta(|x|^{\frac{1}{2n}}\zeta)$ at infinity and such that the set $\{\zeta: \Im g(\zeta)>0\}$ has a convenient topology.

\medskip

We define $g$ in the form
\begin{equation}\label{def g-}
g(\zeta)=\sum_{j=1}^nc_{j}(\zeta^2-\zeta_0^2)^{\frac{2j+1}{2}},
\end{equation}
so that $g(\zeta)$ is analytic in $\mathbb C\setminus [-\zeta_0,\zeta_0]$.

We fix the constants $c_j$ and the branch point $\zeta_0>0$ by
the requirement that
\begin{equation}\label{gtheta-}
|x|^{\frac{2n+1}{2n}}g(\zeta)=\Theta(|x|^{\frac{1}{2n}}\zeta)+\bigO(\frac{1}{\zeta}),\qquad\mbox{ as $\zeta\to\infty$.}
\end{equation}
Expanding this expression as $\zeta\to \infty$ and equating the
coefficients at $\zeta^{2n+1}, \zeta^{2n-1}, \ldots , \zeta^3$
gives (after some calculations which make use of the binomial formula
and induction) that, as $x\to -\infty$,
\begin{align}
&\label{def cn}c_n=\frac{2^{2n}}{2n+1},\\
&\label{def cj}c_{n-m}=\frac{\Gamma(n+\frac{1}{2})}{\Gamma(n-m+\frac{3}{2})}\frac{2^{2n-1}}{m!}\zeta_0^{2m}
+\bigO(|x|^{-\frac{1}{n}}), & m=1, \ldots , n-1.
\end{align}
The coefficient at $\zeta$ yields
\be
\label{zeta0}
\zeta_0^{2n}=\frac{n!\sqrt{\pi}}{2^{2n}\Gamma(n+\frac{1}{2})}+
\bigO(|x|^{-\frac{1}{n}})=\frac{n!^2}{(2n)!}+\bigO(|x|^{-\frac{1}{n}}).
\ee
Here the $\bigO(|x|^{-\frac{1}{n}})$ can also be computed explicitly by taking into account the terms with $\tau_j$ in $\Theta(|x|^{\frac{1}{2n}}\zeta)$.
Note that the rescaling with a factor $|x|^{\frac{1}{2n}}$ in (\ref{def T-}) was necessary to have branch points $\pm\zeta_0$ with a non-zero finite limit as $x\to -\infty$.

We now establish some useful properties of the function $g(\zeta)$.
First, let $\zeta\in(0,\zeta_0)$. Then
\[
|\zeta^2-\zeta_0^2|=\zeta_0^2 z,\quad z\in(0,1).
\]
Note that by (\ref{def cn}), (\ref{def cj}), and (\ref{zeta0}),
\[
c_{n-m}\zeta_0^{2(n-m)}=\frac{n!\sqrt{\pi}}{2\Gamma(n-m+\frac{3}{2})m!}+\bigO(|x|^{-\frac{1}{n}}),
\qquad m=0,\dots,n-1, \quad x\to -\infty.
\]
Thus
\be
g_{\pm}(\zeta)=\frac{n!\sqrt{\pi}}{2}\zeta_0
\sum_{j=1}^n\frac{e^{\pm i\pi(2j+1)/2} z^{j+1/2}}{\Gamma(j+3/2)(n-j)!}
+\bigO(|x|^{-\frac{1}{n}}),
\qquad \zeta\in(0,\zeta_0).
\ee
Therefore,
\be\label{g1}
g_+(\zeta)- g_-(\zeta)=2g_+(\zeta)=
i\sqrt{\pi z}n!\zeta_0\phi(z)+\bigO(|x|^{-\frac{1}{n}}),
\qquad \zeta\in(0,\zeta_0),\quad z\in(0,1),
\ee
where
\be
\phi(z)=\sum_{j=1}^n \frac{(-1)^j z^j}{\Gamma(j+3/2)(n-j)!}.
\ee
We need to determine the sign of $\phi(z)$ for $z\in(0,1)$.
We do it by identifying  $\phi(z)$ with a Jacobi polynomial.
Jacobi polynomials $p^{(\alpha,\beta)}_k(x)$
depend on 2 parameters $\alpha$ and $\beta$, and have
several representations in terms of the hypergeometric function $F$.
We will make use of the following 2 of them:
\begin{eqnarray}
p^{(\alpha,\beta)}_{n-1}(x)=
\frac{\Gamma(n+\alpha)}{(n-1)!\Gamma(1+\alpha)}
F\left(n+\alpha+\beta,-n+1;1+\alpha;{1-x\over 2}\right),\\
p^{(\alpha,\beta)}_{n-1}(x)=
\frac{\Gamma(n+\beta)(x-1)^{n-1}}{2^{n-1}(n-1)!\Gamma(1+\beta)}
F\left(1-n-\alpha,-n+1;1+\beta;{x+1\over x-1}\right).
\end{eqnarray}
From the first of these equations we obtain for $\alpha=3/2$, $\beta=-n-1/2$:
\be\label{r1}
p^{(3/2,-n-1/2)}_{n-1}(1-2z)=
-{\Gamma(n+3/2)\over z}\phi(z).
\ee
On the other hand, the second equation for this particular
Jacobi polynomial yields
\be\label{r2}
p^{(3/2,-n-1/2)}_{n-1}(1-2z)=
{2\Gamma(n+3/2)\over \sqrt{\pi}}
\sum_{m=0}^{n-1}\frac{z^{n-m-1}(1-z)^m}{m!(n-1-m)!(n+1/2-m)},
\ee
which is obviously positive for $z\in(0,1)$.
Therefore, it follows by (\ref{r1}) that
\be
\phi(z)<0,\qquad z\in(0,1),
\ee
and hence, from (\ref{g1}),
\be\label{one}
\Im(g_+-g_-)=2\Im g_+ <0,\qquad \zeta\in(0,\zeta_0),
\ee
for sufficiently large $|x|$ depending on $\tau_j$'s.

If $\zeta\in(-\zeta_0,0)$ the only difference is that $''+''$ and
$''-''$ sides are interchanged and therefore
\be\label{onemore}
\Im(g_+-g_-)=2\Im g_+ >0,\qquad \zeta\in(-\zeta_0,0).
\ee
Moreover, considering the arguments of $\zeta-\zeta_0$ and $\zeta+\zeta_0$,
we easily conclude that
\begin{eqnarray}
\Im g(\zeta)>0,\qquad \zeta\in
\widehat\Upsilon_{1}\cup\widehat\Upsilon_{2n+1}\\
\Im g(\zeta)<0,\qquad \zeta\in
\widehat\Upsilon_{2n+2}\cup\widehat\Upsilon_{4n+2}\label{2}\\
g_+(\zeta)+g_-(\zeta)=0,\qquad \zeta\in(-\zeta_0,\zeta_0).
\end{eqnarray}
Note that the inequalities here and in (\ref{onemore}) hold,
as in (\ref{one}), for sufficiently large $|x|$.

Let us now define
\begin{equation}\label{def S-}
S(\zeta)=
\widehat T(\zeta)\exp\left\{i|x|^{\frac{2n+1}{2n}}g(\zeta)\sigma_3\right\}.
\end{equation}
From the $\Psi$-RH problem we then have

\subsubsection*{RH problem for $S$}

\begin{itemize}
\item[(a)] $S$ is analytic in $\mathbb C\setminus\widehat\Upsilon$.
\item[(b)] $S$ satisfies the following jump
    relations on $\widehat\Upsilon \setminus \{0\}$:
    \begin{align}
        \label{RHP S: b1-}
        S_+(\zeta) &= S_-(\zeta)
            \begin{pmatrix}
                1 & 0 \\
                -i\exp\left\{2i|x|^{\frac{2n+1}{2n}}g(\zeta)\right\} & 1
            \end{pmatrix},
            \qquad\mbox{for $\zeta\in\left(\widehat\Upsilon_1\cup\widehat\Upsilon_{2n+1}\right)\setminus[-\zeta_0,\zeta_0]$,}
        \\[1ex]
        \label{RHP S: b2-}
            S_+(\zeta) &= S_-(\zeta)
            \begin{pmatrix}
                1 & -i\exp\left\{-2i|x|^{\frac{2n+1}{2n}}g(\zeta)\right\}\\
                0 & 1
            \end{pmatrix},
            \qquad \mbox{for $\zeta\in\left(\widehat\Upsilon_{2n+2}\cup\widehat\Upsilon_{4n+2}\right)\setminus[-\zeta_0,\zeta_0]$,}\\[1ex]
        \label{RHP S: b3-}
            S_+(\zeta) &= S_-(\zeta)
            \begin{pmatrix}
                \exp\left\{2i|x|^{\frac{2n+1}{2n}}g_+(\zeta)\right\} & -i \\
                -i & 0
            \end{pmatrix},
            \qquad \mbox{for $\zeta\in(-\zeta_0,0)$,}\\[1ex]
        \label{RHP S: b4-}
            S_+(\zeta) &= S_-(\zeta)
            \begin{pmatrix}
                0 & -i \\
                -i & \exp\left\{-2i|x|^{\frac{2n+1}{2n}}g_+(\zeta)\right\}
            \end{pmatrix},
            \qquad \mbox{for $\zeta\in(0,\zeta_0)$,}
    \end{align}
\item[(c)] At infinity,
    \begin{equation}\label{RHP S: c-}
        S(\zeta)=I+\bigO(\zeta^{-1})),
        \qquad \zeta\to\infty.\end{equation}
\item[(d)] Near the origin,
    \begin{equation}\label{RHP S: d-}
        S(\zeta)=\begin{cases}\begin{array}{ll}\bigO\begin{pmatrix}|\zeta|^{1/2}&|\zeta|^{-1/2}\\ |\zeta|^{1/2}& |\zeta|^{-1/2}\end{pmatrix},&\mbox{ as $\zeta\to 0$, $\Im\zeta>0$,}\\
        \bigO\begin{pmatrix}|\zeta|^{-1/2}&|\zeta|^{1/2}\\ |\zeta|^{-1/2}& |\zeta|^{1/2}\end{pmatrix},&\mbox{ as $\zeta\to 0$, $\Im\zeta<0$.}
        \end{array}\end{cases}
    \end{equation}
\end{itemize}

It follows from (\ref{one})--(\ref{2}) that the jump matrices for $S$ tend exponentially fast to the identity matrix away from the interval $[-\zeta_0,\zeta_0]$.
On $(-\zeta_0,\zeta_0)$, the diagonal entries decay exponentially fast.
The condition (c) is also easy to verify.

\subsubsection{Parametrix outside $\pm\ze_0$}

Ignoring exponentially small jumps and small neighborhoods of the branch points $\pm\ze_0$, we are led to a problem for a parametrix $P^{(\infty)}$:

\subsubsection*{RH problem for $P^{(\infty)}$}
\begin{itemize}
    \item[(a)] $P^{(\infty)}$ is analytic in
$\mathbb{C}\setminus [-\zeta_0, \zeta_0]$.
    \item[(b)] $P^{(\infty)}$ satisfies the following jump relations on
    $(-\zeta_0, 0)\cup (0, \zeta_0)$, with the intervals $(-\zeta_0, 0)$ and $(0, \zeta_0)$ both oriented away from the origin,
    \begin{align}
        P^{(\infty)}_+(\zeta)&=P^{(\infty)}_-(\zeta)
            \begin{pmatrix}
                0 & -i \\
                -i & 0
            \end{pmatrix}.
    \end{align}
    \item[(c)] At infinity,
    \begin{equation}\label{RHP Pinfty:c2}
        P^{(\infty)}(\zeta)=I+\bigO(\zeta^{-1}),\qquad \zeta\to\infty.
    \end{equation}
    \item[(d)] Near the origin,
    \begin{equation}\label{RHP Pinfty: d-}
        P^{(\infty)}(\zeta)=\begin{cases}\begin{array}{ll}\bigO\begin{pmatrix}|\zeta|^{1/2}&|\zeta|^{-1/2}\\ |\zeta|^{1/2}& |\zeta|^{-1/2}\end{pmatrix},&\mbox{ as $\zeta\to 0$, $\Im\zeta>0$,}\\
        \bigO\begin{pmatrix}|\zeta|^{-1/2}&|\zeta|^{1/2}\\ |\zeta|^{-1/2}& |\zeta|^{1/2}\end{pmatrix},&\mbox{ as $\zeta\to 0$, $\Im\zeta<0$.}
        \end{array}\end{cases}
    \end{equation}

\end{itemize}

It is easily seen that this problem has a unique solution which
can be found as follows. First, let $\wt P(\zeta)$ satisfy
the same conditions (a), (b), and (c) as  $P(\zeta)$, and write it in the form
\be
\wt P(\zeta)={1\over 2}  \begin{pmatrix}1&-1\\1&1\end{pmatrix}
L(\zeta)
\begin{pmatrix}1&1\\-1&1\end{pmatrix}.
\ee
Then $L(\zeta)$ has a jump on $(-\zeta_0,\zeta_0)$ with the diagonal
jump matrix $\begin{pmatrix}-i&0\\0&i\end{pmatrix}$, and $L(\infty)=I$.
Therefore, considering the 2 scalar RH problems, we immediately
obtain a solution
\be
L(\zeta)=a(\zeta)^{\sigma_3},\qquad a(\zeta)=
\frac{\zeta^{1/2}}{(\zeta^2-\zeta_0^2)^{1/4}}.
\ee
We assume that the roots are positive on the real axes to the right of the
respective branch points, and the cuts go along the real axis to the left.
Note that the corresponding $\wt P(\zeta)$ satisfies the conditions (a), (b),
and (c), but not (d). Let us multiply it on the left by the following
matrix
\be
A=I+{1\over\zeta}
\begin{pmatrix}\alpha&\beta\\ \gamma &\delta\end{pmatrix}
\ee
and choose $\alpha,\beta,\gamma,\delta$ so that $A \wt P(\zeta)$
satisfies (d), and therefore, since the conditions (a,b,c) are not violated,
gives the solution $P^{(\infty)}$ we are looking for.
A straightforward calculation yields $\alpha=\beta=-i\zeta_0/2$,
$\gamma=\delta=i\zeta_0/2$. Thus, we obtain
\be
P^{(\infty)}(\zeta)=
\left[ I+{i\zeta_0\over 2\zeta}
\begin{pmatrix}-1&-1\\ 1 & 1\end{pmatrix}
\right]
{1\over 2}  \begin{pmatrix}1&-1\\1&1\end{pmatrix}
\left(\frac{\zeta^{1/2}}{(\zeta^2-\zeta_0^2)^{1/4}}\right)^{\sigma_3}
\begin{pmatrix}1&1\\-1&1\end{pmatrix}.
\ee

As $\zeta\to\infty$, we clearly have
\begin{equation}\label{Pinfty-}
P^{(\infty)}(\zeta)=I+\frac{i\zeta_0}{2\zeta}
\begin{pmatrix}-1&-1\\1&1\end{pmatrix}+\bigO(\zeta^{-2}).
\end{equation}
The off-diagonal entries in the coefficient at $\zeta^{-1}$ in this expansion will determine the leading order asymptotics for $q(x)$.

\subsubsection{Local parametrices near $-\zeta_0$ and $\zeta_0$}
In the neighborhoods $U$ of $\zeta_0$ and $\wt U$ of $-\zeta_0$, the
local parametrices are constructed in terms of Airy function.
The analysis is standard, and we only present the result.
In $\wt U$, we have
\begin{multline}
P(\ze)=\hat E(\ze)A\left(|x|^\frac{2n+1}{3n}\lb\right)
\exp\left\{i|x|^{\frac{2n+1}{2n}}g(\zeta)\sigma_3\right\},\\
\hat E(\ze)=P^{(\infty)}(\zeta){1\over 2i}
\begin{pmatrix}1&-i\\1&i\end{pmatrix}
\left(|x|^\frac{2n+1}{3n}\lb\right)^{\sigma_3/4},\qquad
\lb(\ze)^{3/2}={3\over 2}e^{\mp 3\pi i/2}g(\ze),
\end{multline}
where the minus sign is taken in the upper half-plane, and plus, in the lower.
The function $A$ has the following form in the sector between
$\wt\Upsilon_{2n+1}$ and the real-axis-part of the contour:
\be
A(z)=2\sqrt{\pi}
\begin{pmatrix} i \Ai(z) & e^{i\pi/3} \Ai(e^{4\pi i/3}z)\\
{d\over dz}\Ai(z) & -i e^{i\pi/3}{d\over dz} \Ai(e^{4\pi i/3}z)
\end{pmatrix}
\ee
and is constructed in the other sectors using the jump conditions.

We have the crucial properties: $P(\ze)S(\ze)^{-1}$ is analytic in $\wt U$,
and $P(\ze)P^{(\infty)}(\zeta)^{-1}=I+\bigO(|x|^{-\frac{2n+1}{2n}})$
uniformly on $\partial\wt U$ as $|x|\to\infty$.
The parametrix $P(\ze)$ for $\ze\in U$ is obtained using the transformation
$\sigma_1 A(-z)\sigma_1$.

Note that these parametrices do not contribute to the leading-order
asymptotics for $q(x)$.

\subsubsection{Final RH problem and asymptotics for $q$ at $-\infty$}

\begin{figure}[t]
\begin{center}
    \setlength{\unitlength}{1mm}
    \begin{picture}(100,30)(0,0)
        \put(30,15){\thicklines\circle*{.8}} \put(30,15){\circle{15}}
        \put(70,15){\thicklines\circle*{.8}} \put(70,15){\circle{15}}
        \put(71,22){\thicklines\vector(1,0){.0001}}
        \put(31,22){\thicklines\vector(1,0){.0001}}
        \put(37.15,15){\line(1,0){25.7}}
        \put(58,15){\thicklines\vector(1,0){.0001}}
        \put(42,15){\thicklines\vector(-1,0){.0001}}
        \put(76.6,17.9){\line(3,1){25}}
        \put(88.9,22){\thicklines\vector(3,1){.0001}}
        \put(76.6,12.1){\line(3,-1){25}}
        \put(88.9,8){\thicklines\vector(3,-1){.0001}}
        \put(23.4,17.9){\line(-3,1){25}}
        \put(11.1,22){\thicklines\vector(-3,1){.0001}}
        \put(23.4,12.1){\line(-3,-1){25}}
        \put(11.1,8){\thicklines\vector(-3,-1){.0001}}
        \put(50,16){0}
        \put(50,15){\thicklines\circle*{.8}}
        \put(28,16){\small $-\zeta_0$}
        \put(70,16){\small $\zeta_0$}
    \end{picture}
    \caption{Contour for $R$.}
\label{figureSigmaR-3}
\end{center}
\end{figure}
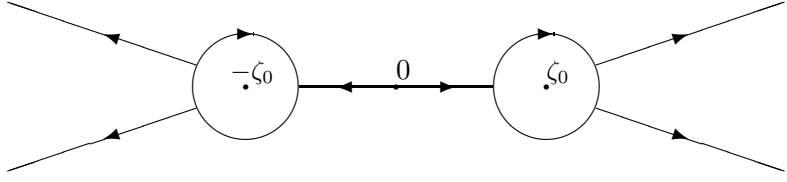
As before we define $R$ by
\begin{align}\label{def R1-3}
&R(\zeta)=S(\zeta)P(\zeta)^{-1}&&\mbox{ for $\zeta\in U\cup\wt U$,}\\
&\label{def R2-3}R(\zeta)=S(\zeta)P^{(\infty)}(\zeta)^{-1},&&
\mbox{ for  $\zeta\in \mathbb C\setminus (U\cup\wt U)$,}
\end{align}
so that $R$ satisfies the following RH problem.
\subsubsection*{RH problem for $R$}
\begin{itemize}
\item[(a)] $R$ is analytic in $\mathbb C\setminus \Sigma_R$, with $\Sigma_R$ as shown in Figure
\ref{figureSigmaR-3}.
\item[(b)] $R$ satisfies the jump relations $R_+=R_-v_R$ on $\Sigma_R$, with
\begin{align}
&\hspace{-2cm}v_R(\zeta)=P(\zeta)P^{(\infty)}(\zeta)^{-1},\quad
\mbox{ for $\zeta\in \partial U\cup\partial\wt U$,}\\
&\hspace{-2cm}v_R(\zeta)=P^{(\infty)}(\zeta)v_S(\zeta)
P^{(\infty)}(\zeta)^{-1},
\quad\mbox{ for $\zeta\in\Sigma_R\setminus
(\overline{U\cup\wt U}\cup [-\ze_0+\ep,\ze_0-\ep])$,}\\
&\hspace{-2cm}v_R(\zeta)=I+P^{(\infty)}_{-}(\zeta)
\begin{pmatrix}
0& i\exp\left\{2i|x|^{\frac{2n+1}{2n}}g_+(\zeta)\right\}\\
0&0\end{pmatrix}
P^{(\infty)}_{-}(\zeta)^{-1},
\quad\mbox{ for $\zeta\in(-\ze_0+\ep,0)$,}\\
&\hspace{-2cm}v_R(\zeta)=I+P^{(\infty)}_{-}(\zeta)
\begin{pmatrix}
0& 0\\
i\exp\left\{-2i|x|^{\frac{2n+1}{2n}}g_+(\zeta)\right\}&0\end{pmatrix}
P^{(\infty)}_{-}(\zeta)^{-1},
\quad\mbox{ for $\zeta\in(0,\ze_0-\ep)$.}
\end{align}
\item[(c)] $R(\zeta)\to I$ as $\zeta\to\infty$,
\end{itemize}
where $\ep$ is the radius of the neighborhoods $U$ and $\wt U$.
Note that, because of $(\ref{RHP S: d-})$ and $(\ref{RHP Pinfty: d-})$, $R$ is bounded near the origin.
From the last sections we recall that
$v_R=I+\bigO(|x|^{-\frac{2n+1}{2n}})$ uniformly on $\Sigma_R$.
It follows in a standard way that $R(\ze)=I+\bigO(|x|^{-\frac{2n+1}{2n}})$
uniformly in $\ze$ as $x\to -\infty$.
Therefore, we obtain from (\ref{def R2-3}), (\ref{Pinfty-}), (\ref{def S-}),
and (\ref{RHP Psi: c}) that
\begin{equation}
\Psi_\infty=\frac{i\ze_0}{2}|x|^{1\over 2n}
\begin{pmatrix}-1&-1\\1&1\end{pmatrix}+\bigO(|x|^{-1}),
\end{equation}
so that, by (\ref{qPsi}),
\begin{equation}\label{q-final}
q(x)=\zeta_0|x|^{\frac{1}{2n}}+\bigO(|x|^{-1}),
\qquad\mbox{ as $x\to -\infty$},
\end{equation}
which, in combination with (\ref{zeta0}),
proves the final part of Theorem \ref{theorem: q}.

For later use, we again need asymptotics for $E$ defined by (\ref{defdefE}).
In Section \ref{section Painleve}
we show that $E(\ze)$ is analytic near the origin,
and we can write
\[
E(\ze)=E_{0}(I+E_{1}\ze+o(\ze)).
\]
A similar analysis to the one following (\ref{defdefE})
for $x\to +\infty$ gives
\begin{equation}\label{as E011-}
E_{0,11}=o(1), \qquad\mbox{ as $x\to -\infty$}.
\end{equation}
and
\be\label{as E121}
E_{1,21}=o(1), \qquad\mbox{ as $x\to -\infty$}.
\ee

\subsubsection{Asymptotics at $-\infty$ for other solutions of the Painlev\'e II hierarchy}


%
%
%
%
%
%

Similarly to the situation at $+\infty$, the asymptotics at $-\infty$ only
do not uniquely fix our special solution of $\P2n$.
In a similar way as was shown for the Painlev\'e II equation itself (see \cite{FIKN}), one has the freedom to choose some of the even Stokes multipliers $s_2, s_4, \ldots , s_{2n}$ different from zero without destroying the above RH analysis.

Under the assumption that the even Stokes multipliers are zero, the method works only if the jump contour connects the branch points $\pm\zeta_0$  with infinity within every Stokes sector corresponding to a non-zero Stokes multiplier.
A simple analysis of $g(\ze)$ shows that this only leads to decaying jump matrices for the Stokes multipliers $s_1$ and $s_{2n+1}$, so the other odd multipliers
should be zero. Recalling Section \ref{section general+},
we can conjecture that the restriction
\[s_2=s_3= \cdots =s_{2n}=0\]
is needed in order to connect the asymptotics of $q(x)$ at $+\infty$
and $-\infty$.
In order to preserve the symmetry relation (\ref{symmetry real}) and thus to have a real solution, the only possibility then is that $s_1=s_{2n+1}=-i$.

\medskip

Let us stress once again that this is not a proof that the
Painlev\'e solution we discuss is uniquely determined by the reality and by its asymptotics. In principle a different asymptotic analysis of the RH problem could still lead to the same asymptotics for $q$. However in view of the direct dependence of the leading order asymptotics for $q$ on the outside parametrix, it would be rather surprising if other solutions with the same asymptotics existed.

\section{Higher order Painlev\'e II formula for the Fredholm determinant}\label{section Painleve}

In this section, we will relate the $X$-RH problem for the
Fredholm determinant to the $\Psi$-RH problem
for the Painlev\'e II hierarchy, namely, to the one corresponding to the special solution of $\P2n$ we described in the previous section. By (\ref{identityFredholmX}), this will enable us to express the logarithmic derivative of $\det(I-K_s^{(k)})$ in terms of the Painlev\'e II hierarchy and to prove Theorem \ref{theorem: TW}.

\subsection{RH problem for the Painlev\'e XXXIV hierarchy and
the Fredholm determinant}\label{section RHP U}

Let us define the following function
$U=U^{(n)}(\zeta;x,\tau_1, \ldots, \tau_{n-1})$ in terms of the solution
$\Psi^{(n)}$ of the RH problem stated in Section \ref{section: RHP}:
\begin{equation}\label{UinPsi}
U^{(n)}(\zeta)=\frac{1}{\sqrt 2}\zeta^{-\frac{1}{4}\sigma_3}\begin{pmatrix}1&1\\-1&1\end{pmatrix}\Psi^{(n)}(i\zeta^{\frac{1}{2}};x,\tau_1,\ldots,\tau_{n-1})
e^{-\frac{1}{4}\pi i\sigma_3}.
\end{equation}
Here we take as usual the branches of the fractional powers which are analytic in $\mathbb C\setminus (-\infty,0]$ and positive on $(0,+\infty)$. 

\medskip

To obtain a RH problem for $U$ from that for $\Psi$, note first
that the $\Psi$-RH problem is defined in the plane of the variable
$u=i\ze^{1/2}$. Because of the symmetry $\sigma_1\Psi(-u)\sigma_1=\Psi(u)$,
it is enough to consider only the half-plane $\Im u\ge 0$. The function
$\ze=(-i u)^2$ maps it onto the whole $\ze$-plane. Now it is easy to verify
that $U$ satisfies the following conditions.

\subsubsection*{RH problem for $U$}
\begin{itemize}
    \item[(a)] $U(\ze)$ is analytic in
$\mathbb{C}\setminus\Sigma$, where the contour $\Sigma$ is
the one shown in Figure \ref{newfig},
    \begin{equation}
    \label{def Sigma2}\Sigma=\Sigma_1\cup\Sigma_2\cup\Sigma_3\cup\{0\}.\end{equation}
    \item[(b)] The boundary values
     of $U$ satisfy the relations:
    \begin{align}
        U_+(\zeta)&=U_-(\zeta)
            \begin{pmatrix}
                1 & 0 \\
                1 & 1
            \end{pmatrix},&& \mbox{for $\zeta\in\Sigma_1\cup\Sigma_3$,}\\[1ex]
        U_+(\zeta)&=U_-(\zeta)
            \begin{pmatrix}
                0 & 1 \\
                -1 & 0
            \end{pmatrix},&& \mbox{for
            $\zeta\in\Sigma_2$.}
    \end{align}
    \item[(c)] As
    $\zeta\to\infty$,
    \begin{equation}\label{RHP U:c}
        U(\zeta)=\zeta^{-\frac{1}{4}\sigma_3}N\left(I+\bigO(\zeta^{-\frac{1}{2}})\right)
        e^{-i\Theta(i\zeta^{\frac{1}{2}})\sigma_3},
    \end{equation}
    with $N$ given by (\ref{def: N}) and $\Theta$ by (\ref{def Theta}).
    \item[(d)] As $\zeta\to 0$,
    \begin{equation}\label{RHP U:d}
        U(\zeta)=\bigO(|\zeta|^{-1/2}).
    \end{equation}
\end{itemize}

Although we do not use this fact, it is worth noting that the RH problem for $U$ is the RH problem associated to one solution of the $2n$-th order equation in the Painlev\'e XXXIV hierarchy.

Our aim is to identify $U=U^{(n)}$ with the RH problem
for $X^{(k)}$ from Section \ref{section fredholm} if $n=2k+1$.
Let us first identify the parameters $(s,t_0, \ldots, t_{2k-1})$ for $X$ with the parameters $(x,\tau_1, \ldots, \tau_{2k})$ for $U$. We do this by requiring that
\begin{equation}\label{identificationt}
\theta(\zeta+s;t_0, \ldots, t_{2k-1})=i\Theta(i(c\zeta)^{1/2})+\bigO(\frac{1}{\zeta}),\qquad\mbox{ as $\zeta\to\infty$,} \qquad c=4^{-\frac{4k+1}{4k+3}},
\end{equation}
where $\theta$ is defined in (\ref{def theta}), and $\Theta$ in
(\ref{def hattheta}). The value of $c$ follows from comparison of
the leading order terms in $\zeta$ for $\theta$ and $\Theta$.
Expanding $\theta(\zeta+s)$ for $\zeta\to\infty$
similarly as in (\ref{expansion theta})-(\ref{def aj1}) for fixed $s$, we find
\begin{equation}\label{expansion theta2}
\theta(\zeta+s)=\sum_{j=0}^{2k+1} b_j \zeta^{\frac{2j+1}{2}}+\bigO(\zeta^{-1/2}),
\end{equation}
with
\begin{multline}
\label{def bj1}
b_j(s)=\frac{\Gamma(2k +\frac{3}{2})}{\Gamma(j+\frac{3}{2})\Gamma(2k+2-j)}s^{2k+1-j}\\
-\sum_{\ell=j}^{2k-1}(-1)^\ell t_\ell \frac{\Gamma(\ell +\frac{1}{2})}{\Gamma(j+\frac{3}{2})\Gamma(\ell - j+1)}s^{\ell-j}.
\end{multline}
Comparing the same powers of $\zeta$ in (\ref{identificationt}) we obtain that $x=x(s;t_0, \ldots, t_{2k-1})$ and $\tau_j=\tau_j(s;t_0, \ldots, t_{2k-1})$ are given as follows:
\begin{align}
&\label{x}x(s)=-c^{-1/2}b_0(s;t_0, \ldots, t_{2k-1}),\\
&\label{ttau}\tau_{j}(s)=(2j+1)4^{-j}c^{-j-\frac{1}{2}}b_j(s;t_0, \ldots, t_{2k-1}), &j=1, \ldots , 2k.
\end{align}
Let us now set
\begin{equation}
\widehat X(\zeta;t_0, \ldots, t_{2k-1})=c^{\frac{1}{4}\sigma_3}U(c(\zeta-s);x,\tau_1,\ldots, \tau_{2k}),
\end{equation}
with $x$ and $\tau_j$ given by (\ref{x}) and (\ref{ttau}).
It is easy to verify that $\widehat X$ satisfies the jump conditions (\ref{RHP X:b1})--(\ref{RHP X:b2}) of the RH problem for $X$,
and therefore  $\widehat XX^{-1}$ is a holomorphic function in
$\mathbb C\setminus\{0\}$ for any RH solution $X$.
Using (\ref{identificationt}) and the asymptotic behavior of $U$, we see that also the asymptotic condition (\ref{RHP X:c}) holds for $\widehat X$
and $\widehat XX^{-1}$ is bounded at infinity.
By (\ref{RHP X:d}) and (\ref{RHP U:d}), $\widehat XX^{-1}$ can only have a removable singularity at $0$. Therefore,
by Liouville's theorem, $\widehat XX^{-1}$ is a constant matrix, and so
$\widehat X$ solves the RH problem for $X$. Since (\ref{identityFredholmX})
holds for any solution of the $X$-RH problem, we obtain
\begin{equation}\label{FU}
F(s)=\frac{d}{ds}\ln\det(I-K_s^{(k)})=\frac{4^{-\frac{4k+1}{4k+3}}}{2\pi i}
\left.\left(U^{-1}(z)U_z(z)\right)_{21}\right|_{z\searrow 0}.
\end{equation}

\subsection{Proof of Theorem \ref{theorem: TW}}
In order to prove Theorem \ref{theorem: TW}, we will consider the
$x$-derivative of $U$.
As already mentioned, the RH problem for $U$ (see Section \ref{section RHP U}) is not uniquely solvable but has a family of solutions of the form
\begin{equation}
\widehat U(\zeta)=\begin{pmatrix}1&0\\ \eta&1\end{pmatrix}U(\zeta),
\end{equation}
where $U$ is defined by (\ref{UinPsi}), and $\eta$ is independent of $\zeta$ but may depend on $x, \tau_1, \ldots , \tau_{2k}$. Although
the identity
\begin{equation}\label{FUhat}
F(s)=\frac{d}{ds}\ln\det(I-K_s^{(k)})=\frac{4^{-\frac{4k+1}{4k+3}}}{2\pi i}
\left.\left(\widehat U^{-1}(\ze)\widehat U_\zeta(\ze)\right)_{21}
\right|_{\zeta\searrow 0},
\end{equation}
which we obtained in (\ref{FU}), is independent of the choice of $\eta$, the linear system satisfied by $\widehat U$ does depend on $\eta$.
A convenient choice for us is $\eta=-q$. Then we obtain from (\ref{UinPsi}) and (\ref{Lax PII}), (\ref{B}) that $U$ satisfies the following differential equation with respect to $x$ (which is one of the equations in the standard Lax pair for Painlev\'e XXXIV):
\begin{equation}\label{diff U x}
\widehat U_x(\zeta;x)=\begin{pmatrix}0&-1\\-\zeta-(q_x+q^2)&0\end{pmatrix}
\widehat U(\zeta;x).
\end{equation}

Let us define a function $E$ by the equation
\begin{equation}\label{hat U E}
\widehat U(\zeta)=E(\zeta)\begin{pmatrix}1&\frac{1}{2\pi i}\ln
\zeta\\0&1\end{pmatrix}C(\zeta),
\end{equation}
where
\begin{equation}\label{def C2}
C(\zeta)=\begin{cases}\begin{array}{ll} I,&\mbox{ in region {\rm I},}\\
\begin{pmatrix}
1&0\\-1&1
\end{pmatrix},&\mbox{ in region {\rm II},}\\
\begin{pmatrix}
1&0\\1&1
\end{pmatrix},&\mbox{ in region {\rm III}.}
\end{array}
\end{cases}
\end{equation}
The sectors I, II, and III are those indicated in Figure \ref{figure: Sigma1}
if $s=0$.
Considering the RH problem for $U$ and
choosing the branch of the logarithm with the cut along $(-\infty,0)$, it
is easy to verify that (\ref{hat U E}) defines an analytic $E(\ze)$ in
a neighborhood of zero with $\det E=1$.

Writing a Taylor series for $E$ near 0,
\begin{equation}
E(\zeta;x)=E_0\left(I+E_1\zeta+\bigO(\zeta^2)\right), \qquad \mbox{ as $\zeta\to 0$},
\end{equation}
we obtain from (\ref{FUhat}) that
\begin{equation}\label{FE}
\frac{d}{ds}\ln\det(I-K_s^{(k)})=Q(x;\tau_1, \ldots , \tau_{2k}),
\end{equation}
with
\begin{equation}\label{def Q}
Q(x;\tau_1, \ldots , \tau_{2k})=\frac{4^{-\frac{4k+1}{4k+3}}}{2\pi i}
E_{1,21}\left(x;\tau_1, \ldots, \tau_{2k}\right).
\end{equation}
Using (\ref{diff U x}) we also find
\begin{align}
&\label{diff E0}E_{0,x}=\begin{pmatrix}0&-1\\-(q_x+q^2)&0\end{pmatrix}E_0,\\
&E_{1,x}=E_0^{-1}\begin{pmatrix}0&0\\-1&0\end{pmatrix}E_0,
\end{align}
and thus
\begin{equation}
E_{1,21,x}=-E_{0,11}^2.
\end{equation}
In other words,
\begin{equation}\label{Fredholmfinal}
Q'(x)=u(x)^2, \qquad u(x)=\frac{e^{\frac{i\pi}{4}}}{\sqrt{2\pi}}2^{-\frac{4k+1}{4k+3}}E_{0,11}(x).
\end{equation}

We see from (\ref{as E121}) that $E_{1,21}(-\infty)=0$.
Therefore equations (\ref{def Q}) and
(\ref{Fredholmfinal}) imply the formula (\ref{Qinfty}).

Using (\ref{diff E0}) we obtain the following second-order differential
equation for $u(x)$:
\begin{equation}\label{Fredholmfinal1}
u''(x)=[q_x(x)+q^2(x)]u(x).
\end{equation}
Substituting the asymptotics (\ref{as E011+}) and (\ref{as E011-})
into (\ref{Fredholmfinal}), we obtain the boundary conditions
(\ref{boundary1})-(\ref{boundary2}) for $u$. They determine $u(x)$ uniquely
for a given $q(x)$.
Now it is easy to check directly using
Theorem \ref{theorem: q} that $u(x)$ given by
(\ref{def u0}) is the solution of  (\ref{Fredholmfinal1}) satisfying
these boundary conditions.
This completes the proof of Theorem \ref{theorem: TW}.

\section{The constant problem}\label{const}
In this last section we discuss the multiplicative constant in the
asymptotics of $\det(I-K_{s}^{(k)})$ as $s\to-\infty$.
In contrast to the sine, Airy and Bessel kernel
determinants, we do not provide a compact expression for this constant in
terms of single integrals of elementary functions.
For this reason and for simplicity of the argument,
we assume that $k=1$ and $t_0=t_1=0$ for the rest of this section
(a generalization being straightforward).
In this case, the expression (\ref{k1}) has the form
\be\label{kk1}
\ln \det(I-K_s^{(1)})=-\frac{5^2}{2^8\cdot 7}|s|^7
+\chi^{(1)}-\frac{3}{8}\ln|s|+\bigO(|s|^{-2}),\qquad s\to-\infty,
\ee
where $\chi^{(1)}$ is the logarithm of the constant in question.
Comparing this expression with Theorem \ref{theorem: TW}
for $k=1$, we obtain
a representation of the constant in terms of our special solution $q(x)$
of ${\rm P}_{{\rm II}}^{(3)}$:
\begin{multline}\label{c1}
\chi^{(1)}=\lim_{s\to-\infty}
\left(
\frac{5^2}{2^8\cdot 7}|s|^7+\frac{3}{8}\ln|s|
\right.
\\
\left.
-
\int_{s}^{+\infty}Q\left[-2^{5/7-3}5t^3;2^{1/7-2}15t^2,2^{-3/7}5t\right] dt
\right),
\end{multline}
which is basically (using (\ref{Qinfty}) and (\ref{def u0})) a closed expression for $\chi^{(1)}$ in terms of the
Painlev\'e II solution $q$ with $k=1$.

This is a generalization of a similar formula, expressing $\chi^{(0)}$ in terms
of ${\rm P}_{{\rm II}}^{(1)}$, in the case of the Airy-kernel determinant.

Let us now try to generalize the argument in \cite{DIK} which led to the
simple expression (\ref{chi}) for $\chi^{(0)}$.
Consider the kernel (\ref{kernel}) with $V(x)$ given by
\[
V(x)=4\left(\frac{1}{5}x^4-\frac{4}{3}x^3+3x^2-2x\right)
\]
for the polynomials $p_k(x)$ orthogonal with the weight
$e^{-nV(x)}$ on the half-axis $(0,+\infty)$ (we choose a half-axis instead of
$\mathbb R$, and hence this $V$ instead of (\ref{V}),
for convenience only). The corresponding mean eigenvalue density
is supported on $[0,2]$ with the $k=1$ type singularity at $x=2$:
\[
\psi_V(x)=\frac{8}{5\pi}x^{1/2}(2-x)^{5/2}\chi_{[0,2]}.
\]
\begin{remark}
Note that, on the half-axis, one can find a potential $V$ of degree $3$ with a singular endpoint. However, this does not simplify the argument below.
\end{remark}
We now expect equation (\ref{distr}) to hold at the endpoint $b=2$.
To prove the convergence of the determinants (\ref{distr}), one would need to
obtain global estimates for the polynomials $p_k(x)$ (cf. \cite{DIK, DG}), which
should be possible in view of \cite{CV2}.
Note that (\ref{distr}) for
our $V(x)=\wt V(x)$ is equivalent to the following:
\be
\lim_{n\to\infty}
\frac{D_n(2+s/(cn^{2/7}))}{D_n(+\infty)}=\det(I-K_s^{(1)}),
\ee
where
\be
D_n(\al)={1\over n!}\int_0^\al\cdots\int_0^\al
\prod_{0\le i<j\le n-1}(x_i-x_j)^2
\prod_{j=0}^{n-1}e^{-nV(x_j)}dx_j.
\ee

In view of (\ref{kk1}), this formula would provide a representation
for $\chi^{(1)}$ in terms of a limit of multiple integrals.
It can be however simplified following \cite{DIK}.
Namely, there exists a differential identity for
${d\over d\al}\ln D_n(\al)$ in terms of the polynomials orthogonal with
the weight $e^{nV(x)}$ on the interval $(0,\al)$. Analysing the
related RH problem, one may be able to obtain a uniform asymptotics for
the polynomials, and hence, for ${d\over d\al}\ln D_n(\al)$ in the range
$\al\in (0,2+s/(cn^{2/7}))$, $s<s_0<0$ for a sufficiently large $|s_0|$,
$n>|s_0|$. On the other hand, it is easy
to obtain a series expansion for $D_n(\al)$ for $n$ fixed and $\al\to 0$.
Combining these results, one could integrate ${d\over d\al}\ln D_n(\al)$
from $\al\to 0$ to $2+s/(cn^{2/7})$, and therefore one may be able to obtain,
as in \cite{DIK}, an expansion for $D_n(2+s/(cn^{2/7}))$ as $n\to\infty$
(and $|s|$ sufficiently large) in terms of elementary functions.
Thus, the only non-elementary object to enter the
expression for $\det(I-K_s^{(1)})$, $s<s_0$, and hence the expression for
$\chi^{(1)}$, would be $D_n(+\infty)$. This would appear to be a final
formula for $\chi^{(1)}$. Indeed, note that in \cite{DIK}, for $\chi^{(0)}$,
the corresponding $D_n(+\infty)$ was a Selberg integral and therefore
the corresponding formulas simplified to (\ref{chi}).
In the present case, however, there is no known expression for
$D_n(+\infty)$ in terms of elementary functions (or a fixed number
of integrals thereof). An attempt to derive the asymptotics of
$D_n(+\infty)$ as $n\to\infty$ by continuing the integration of the differential
identity beyond the endpoint $2$ of the measure
is likely to produce an expansion involving integrals of Painlev\'e functions.

\section*{Acknowledgements}
Tom Claeys is a Postdoctoral Fellow of the Fund for Scientific Research - Flanders (Belgium), and was also supported by
Belgian Interuniversity Attraction Pole P06/02, FWO-Flanders project G042709, K.U.Leuven research grant OT/08/33, and by ESF program MISGAM.
Alexander Its was supported
in part by NSF grant \#DMS-0701768 and
EPSRC grant \#EP/F014198/1. Igor Krasovsky was
supported in part by EPSRC grants \#EP/E022928/1
and \#EP/F014198/1.

\end{document}